\documentclass[english,11pt]{article}
\usepackage{xcolor}
\usepackage[skins,breakable]{tcolorbox}
\usepackage{algorithm}
\usepackage{arxiv}

\usepackage{amsmath,amsfonts,bm}

\def\eqref#1{equation~\ref{#1}}

\def\1{\bm{1}}

\def\eps{{\varepsilon}}

\DeclareMathAlphabet{\mathsfit}{\encodingdefault}{\sfdefault}{m}{sl}
\SetMathAlphabet{\mathsfit}{bold}{\encodingdefault}{\sfdefault}{bx}{n}

\def\gG{{\mathcal{G}}}

\def\gK{{\mathcal{K}}}

\def\gY{{\mathcal{Y}}}

\newcommand{\E}{\mathbb{E}}

\newcommand{\sigmoid}{\sigma}

\newcommand{\Var}{\mathrm{Var}}

\newcommand{\Cov}{\mathrm{Cov}}

\usepackage{hyperref}
\usepackage{etoc}
\usepackage{graphicx}
\usepackage{url}
\usepackage{xurl}

\usepackage{algpseudocode}
\usepackage{amsthm}
\newtheorem{proposition}{Proposition}

\newtheorem{fact}{Fact}

\newtheorem{definition}{Definition}
\newtheorem{assumption}{Assumption}
\newtheorem{lemma}{Lemma}

\newcommand{\acc}{\mathsf{acc}}
\newcommand{\LM}{\mathrm{LM}}

\newcommand{\Ber}{\mathrm{Ber}}

\newcommand{\RR}{\mathrm{RR}}

\title{\LARGE Reconciling Privacy with Provenance \\ for Language Model Watermarks}

\author{
  Elise L. Han and Tatsunori Hashimoto \\[0.3em]
  Department of Computer Science\\[0.3em]
  Stanford University
}
\date{}

\begin{document}
\etocdepthtag.toc{main}
\maketitle

\renewcommand{\thefootnote}{}
\footnotetext{Correspondence to: elisehan@stanford.edu. Code is available at \url{github.com/eliselyhan/dp-watermarks}.}
\begin{abstract}
Watermarking has great potential for regulating the use of language models (LMs), but its deployment has been marred by controversy, as many users argue that fine-grained tracking of LM usage violates their privacy. This tension reduces the effectiveness of watermarks, as users engage in evasion strategies such as paraphrasing that can remove much of the watermark signal. We show that provenance and privacy can be reconciled if we shift the goal of watermark design from tracking individuals to estimating aggregate usage. We develop locally differentially private watermarks which provably limit the evidence that any individual text can contribute to detection, while still enabling accurate estimation of population frequencies. We introduce the Red-Green Sequence (RGS) watermark, a private and computationally efficient watermark that achieves accurate frequency estimation at high privacy levels across various models, sample sizes and generation entropies in settings that reflect realistic deployment scenarios. On C4 language modeling tasks, we attain $0.0019$ MSE with
$0.0038$ mean absolute bias at a low $\varepsilon$ of 1, while in more challenging settings such as an essay writing task with only 100 text samples, we still find errors on the order of $0.0113$ MSE with $0.004$ mean absolute bias at $\varepsilon=2$. \end{abstract}

\section{Introduction}

Watermarking has emerged as a promising approach to enable content provenance through injecting recoverable signals into each piece of model generated text \citep{pmlr-v202-kirchenbauer23a, kuditipudi2024robustdistortionfreewatermarkslanguage}. Among provenance techniques, watermarking is unique in providing a false positive guarantee. If the detection key is kept secret, the false positive rate of identifying a piece of text as being LM generated can be precisely controlled via the generation mechanism \citep{zhao2023provablerobustwatermarkingaigenerated}.

While provable guarantees for tracking individual texts are powerful for the targeted monitoring and regulation of AI-generated content, they have also faced significant pushback from users in recent times \citep{Spirlet,Haynes-backlash}. A major deployment by Anthropic \citep{claude} has led to accusations of surveillance and concentration of power \citep{Brickner-Wood_2026,Brown_2026}, amidst increasing concerns that steganographic techniques may be used to spy upon users \citep{SitePoint,spymarks}.

The tension between the desire for accurate tracking by regulators and users' desire for privacy has also led to adversarial dynamics \citep{liu-etal-2026-position}, where users actively modify their text to evade watermark detection. Even simple strategies, like paraphrasing by another language model, can erase most of the signal from the watermarks \citep{paraphrase,zhang2025watermarkssandimpossibilitystrong}. This leads to a paradoxical situation where the precision and accuracy of a watermark lead to user behavior that sabotages watermark efficacy. 

Therefore, we ask if we can achieve many of the provenance goals of watermarking without violating individual privacy. We answer the question in the affirmative: we observe that many of the regulatory and monitoring goals of watermarking only require population level tracking \citep{liang2026monitoringaimodifiedcontentscale,delve-into-llm-assisted-writing,ai-collapse-recursive}. In these cases, it is possible to design privacy-preserving watermarks that enable accurate estimation of the prevalence while protecting individuals.

We develop the Red-Green Sequence (RGS) watermark, which is a locally differentially private watermark that provably limits the evidence that any individual text can contribute to detection \citep{DBLP:conf/tcc/DworkMNS06,privacybook}, while enabling efficient estimation of the prevalence of watermarked texts in a population of texts. 

Our watermark builds upon standard red-green style language model watermarks \citep{pmlr-v202-kirchenbauer23a}, but carefully modifies them to be effective in the local differential privacy setting. We find that sequence-level watermarks, which are unappealing in the nonprivate setting, are highly efficient in injecting watermarks in private, low-$\eps$ settings. 

To obtain good estimates of aggregate watermark frequencies from these watermarks, we identify several interesting statistical problems in the bias-variance trade-offs of moment estimators for privacy-preserving watermarking, where the natural variability of watermark entropy across texts leads to different sampling and estimation choices compared to standard, randomized-response style estimators.

We implement this watermark on a range of scenarios, finding our frequency estimates to be highly accurate across a range of settings. In a standard C4 language modeling task, we can attain a low MSE of $0.0019$ with $0.0038$ mean absolute bias at a low $\eps$ of 1, while in more challenging settings like an essay writing task with very few text samples, we still find errors on the order of $0.0113$ MSE with $0.004$ mean absolute bias at $\eps=2$. These results suggest that privacy-preserving watermarking is possible and effective.

\section{Related Work}
Language model watermarks inject recoverable statistical signal into the generated text to enable detection of the presence of the watermark from the text. Prominent token-level watermarks work by reweighting token-level probabilities with a secret key \citep{pmlr-v202-kirchenbauer23a,lee-etal-2024-wrote,liu2024adaptivetextwatermarklarge} or coupling sampling with key randomness \citep{Aaronson_2022, kuditipudi2024robustdistortionfreewatermarkslanguage,li2026textgrain,Dathathri2024ScalableWF}. Among these, the closest to our work is \citet{pmlr-v202-kirchenbauer23a}, who use the idea of slightly upweighting pseudorandomly chosen ``green'' tokens that can later be detected. Our approach similarly uses a colored biasing scheme, but our goals and techniques differ significantly from theirs in that we aim to precisely control the signal rather than maximize it. 

Our watermark preserves privacy by making the key-conditional watermarked text distribution close to the one without the watermark. While distortion-free watermarks \citep{kuditipudi2024robustdistortionfreewatermarkslanguage,hu2024unbiased,wu2024resilientaccessibledistributionpreservingwatermark} achieve a similar goal, their watermarks only preserve the base model distribution in expectation over the sampling randomness in the secret key.  Such equality in marginal distributions does not imply that watermarked and unwatermarked text distributions remain close conditional on a fixed key. In fact, the separation in conditional distributions is what enables high-power detection, which is at odds with privacy. Another related objective is cryptographic undetectability \citep{pmlr-v247-christ24a,christ_pseudorandom_2024}, which ensures the marginal distributions of watermarked and unwatermarked texts are computationally indistinguishable to anyone without the secret key, but not conditional on the key. While \citet{kim2026simplemarkfastmultibittext} constructs watermarks under key-conditional $f$-divergence bounds, which can cover differential privacy constraints, their goal is to recover multi-bit messages from individual texts rather than estimate population watermark prevalence. On the other hand, \citet{li2026optimalestimationwatermarkproportions} estimates the proportion of watermarked content within individual texts rather than across a population of texts, and does not provide privacy guarantees. 

Our watermark guarantees privacy formally through local differential privacy (DP) \citep{DBLP:conf/tcc/DworkMNS06, privacybook}, and the watermark method is inspired by randomized response \citep{Warner1965RandomizedRA}, which enables efficient estimation of the prevalence of sensitive attributes while protecting individual respondents, achieving local DP guarantees. The privacy–utility tradeoff under local DP has been studied by \citet{NIPS2013_5807a685,duchi2013privacy,pmlr-v37-kairouz15}, who establish information-theoretic optimality of randomized response in several settings. Related variants of randomized response have also been used for frequency estimation in a fixed population under local DP \citep{ldp-histograms,ldp-frequency-estimation}. Our setting departs from randomized response and generic local DP estimation by incorporating two important constraints. First, our private mechanism must enable signal to be recoverable from the output text alone, without requiring a separate report from each individual. Second, while randomized response allows for modification of distributions for groups with and without the sensitive attribute, our watermark must modify only the text distributions of the watermarked group, as the unwatermarked group's generation process is beyond watermark control.

\section{Problem Setup}
\subsection{Watermarks for language models}
The process of watermarking involves modifying the standard language model decoding process using a secret key $k$, so that correlations between the choice of the key and text output can be used to detect whether a piece of text was generated by an LM decoder that uses $k$.

More formally, let $x$ be a prompt, and $p_x$ be the base model's distribution over sequences given $x$. We sometimes suppress the dependence on $x$. A watermark consists of a decoder $\Gamma$, which is an algorithm that maps a secret key $k \in \gK$ and distribution $p_x$ to a sequence $y \in \gY$ stochastically, i.e. it specifies a new distribution over completions, usually different from the base model distribution $p_x$, and samples a sequence $y$ from the new distribution. The decoder is designed to allow detectors to detect the presence of the watermark in a text sequence $y$ given the secret key $k$.

For example, suppose $\gY$ consists of only single-token sequences. Then, the Soft Red List watermark of \citet{pmlr-v202-kirchenbauer23a} has the following decoder:
\begin{equation}
\label{eq:kgw-watermark}
\begin{split}
    \Pr[\Gamma(k; p_x)=y \mid k] = \frac{p_x(y) \exp(\delta \1\{y \in \gG\})}{\sum_{y' \in \gY} p_x(y') \exp(\delta \1\{y' \in \gG\})},
\end{split}
\end{equation}
where $\gG$ is the set of green tokens obtained by partitioning the token space into green and red tokens via a hash function with key $k$, and $\delta$ is a parameter that controls the strength of the watermark. The decoder biases the watermark towards outputting more green sequences than usual. 
Generalizing to longer sequences, the watermark applies the same bias towards green tokens at each token position, so a text with many green tokens is strong evidence for the watermark. Consequently, an LM user who is unwilling to disclose their LM usage might paraphrase the watermarked text before publishing, in the hope of reducing the number of green tokens and evading detection. 

Importantly, watermarks are often designed to allow for high-powered detection from a single piece of text. In the case of a red-green list watermark, any moderately long piece of text will carry an exponentially increasing watermark signal, and many watermarking papers show extremely low p-values for high-entropy sequences \citep{pmlr-v202-kirchenbauer23a, kuditipudi2024robustdistortionfreewatermarkslanguage}. 

The core claim of our work is that we can instead consider watermarks that more precisely control the watermarking signal, where individual texts still contribute enough signal to accurately estimate the aggregate prevalence of watermarked texts in a population, but are too weak to reliably identify any individual. To illustrate how we construct this estimator, we first illustrate a similar tension in survey design and how it can be addressed.

\subsection{From individual detection to aggregate estimation: randomized response}
\label{sec:rr}
Consider the scenario of reporting a sensitive attribute in response to a survey question, such as ``have you ever used illegal drugs?'' It has long been known that survey respondents can often alter their answers to avoid the possibility that their true, sensitive status may later be revealed \citep{sensitive-surveys}. To address this, survey designers have developed techniques that incentivize respondents to answer truthfully while maintaining the integrity of the survey \citep{rr-meta-analysis}. One such technique is randomized response (RR) \citep{Warner1965RandomizedRA}, which allows respondents to reveal sensitive information to interviewers with plausible deniability.

Consider a fixed pool of $n$ respondents whose sensitive attributes
are $W_1,\ldots,W_n\in\{0,1\}$, with $W_i=1$ indicating that
respondent $i$ has the sensitive attribute. Suppose that the true prevalence of the sensitive attribute in the pool is $\lambda = \frac1n \sum_{i=1}^n W_i$. 

In a generalized randomized response (RR) mechanism, each
respondent engages independently with a randomization device with $0<\pi_0<\pi_1<1$, and outputs a noisy answer $H_i \in \{0, 1\}$ with these probabilities:
\begin{equation}
\label{eq:rr-probabilities}
\begin{aligned}
    &\Pr[H_i=1\mid W_i=1]=\pi_1,
    &\qquad &\Pr[H_i=1\mid W_i=0]=\pi_0, \\
    &\Pr[H_i=0\mid W_i=1]=1-\pi_1,
    &\qquad &\Pr[H_i=0\mid W_i=0]=1-\pi_0.
\end{aligned}
\end{equation}
Standard symmetric RR sets $\pi_1=1-\pi_0$. RR provides plausible deniability to the respondents: if they had reported a sensitive answer $H=1$, they can still claim that it was due to randomization since there is $\pi_0 > 0$ probability that they are forced to report $H=1$ even if their true status is $W=0$.

Such plausible deniability can be formalized by the notion of local differential privacy (DP) \citep{DBLP:conf/tcc/DworkMNS06,privacybook}. Consider a randomized mechanism which maps a private bit $W$ to a discrete output $Y \in \gY$. The mechanism is $\eps$-locally differentially private \citep{privacybook} if 
\begin{equation}
\label{eq:local-dp}
    \Pr[Y=y\mid W=w]
    \leq e^\eps\Pr[Y=y\mid W=w']
    \quad\text{for all }y\in\gY
    \text{ and }w,w'\in\{0,1\}.
\end{equation} 
This condition bounds the likelihood ratio of observing any output under the two possible values of the sensitive bit, thus limiting the evidence that the output can reveal about the sensitive bit. The reporting mechanism in \eqref{eq:rr-probabilities} satisfies local $\eps_{\RR}$-DP, where the privacy parameter
\[ \eps_{\mathrm{RR}} = \max\left\{ \ln\frac{\pi_1}{\pi_0}, \ln\frac{1-\pi_0}{1-\pi_1} \right\}.
\]

While privacy is preserved by adding noise to each person's true bit, aggregating the noisy bits still allows us to recover information about the mean of the true bits in the pool of $n$ respondents, $\lambda =\frac{1}{n} \sum_{i=1}^n W_i$. This is because randomization is precisely controlled by the probabilities $\pi_0, \pi_1$. Writing $\bar{H} = \frac1n \sum_{i=1}^n H_i$, we observe that $\E[\bar{H} \mid W] =  \lambda \pi_1 + (1-\lambda) \pi_0.$
Correspondingly, the moment estimator
\begin{equation}
\label{eq:rr-mom-estimator}
\hat \lambda_{\mathrm{RR}} = \frac{\bar H -\pi_0}{\pi_1 - \pi_0}
\end{equation}
is unbiased and consistent for $\lambda$ \citep{Warner1965RandomizedRA}.

Increasing $\pi_1 - \pi_0$ widens the separation between the reporting probabilities with private bits $W=0$ and $W=1$, which makes each noisy output more informative about the private bit, which increases estimation accuracy but reduces privacy. For binary attributes, it has been shown that the randomized response mechanism with the following reporting probabilities attains optimal information utility under $\eps$-local DP \citep{DBLP:journals/corr/KairouzOV14}: 
\[
\pi_0^*=\sigma(-\eps)=\frac{1}{1+e^\eps}, \qquad \pi_1^*=\sigma(\eps)=\frac{e^\eps}{1+e^\eps}.
\]
This choice maximizes the separation $\pi_1-\pi_0$ and minimizes the variance of the moment estimator in \eqref{eq:rr-mom-estimator} among all binary $\eps$-local DP mechanisms, which also minimizes the MSE since there is no bias.

\subsection{Locally Differentially Private Watermarks}
We have illustrated how randomized response can preserve individual privacy and produce efficient estimation of aggregate prevalence in the survey setting. Inspired by this, we design a watermark that is locally differentially private, and an efficient estimator for the prevalence of watermarked texts in a mixture of texts. First, we formalize the notion of local differential privacy for watermarks by directly adapting \eqref{eq:local-dp}:
\begin{definition}
\label{def:local-dp-watermark}
For some $\eps>0$, a watermark decoder $\Gamma$ is $\eps$-locally differentially private with respect to the base model distribution $p$, if for any key $k$ and any $y \in \gY$:
\begin{equation}
    e^{-\eps} p(y) \leq \Pr[\Gamma(k; p)=y \mid k] \leq e^{\eps} p(y)
\end{equation}
\end{definition}

Denote $p(\cdot)$ as the prior distribution of keys, and 
$p(\cdot|y)$ the posterior distribution of keys given observed text $y$. Let $q(y \mid k)$ be the shorthand for $\Pr[\Gamma(k; p)=y \mid k]$. Definition \ref{def:local-dp-watermark} then implies:
\begin{align*}
p(k\mid y)
&= \frac{q(y\mid k)p(k)}
{\sum_{k'}q(y\mid k')p(k')} \leq \frac{e^\eps p(y)p(k)}
{\sum_{k'}e^{-\eps}p(y)p(k')} = e^{2\eps}p(k).
\end{align*}
Applying the same argument in the other direction gives $p(k\mid y) \geq e^{-2\eps}p(k)$. Viewing the watermark as introducing an association of the text with the secret key \citep{statistical-watermarking-framework}, this condition limits the strength of such association that can be inferred by an observer with the key, and thus provides a natural notion of privacy for arbitrary distributions of non-watermarked text.

\section{The Red-Green-Sequence (RGS) Watermark}

We now present a simple construction for a watermark that satisfies $\eps$-local DP, which we call the Red-Green-Sequence (RGS). RGS closely follows the classic token-level red-green list construction of \citet{pmlr-v202-kirchenbauer23a}, but involves a few subtle changes that arise from our desire to fulfill local differential privacy.

\subsection{Algorithm}
\label{sec:algorithm}
The biggest difference between RGS and \citet{pmlr-v202-kirchenbauer23a} is that RGS operates at the sequence level, i.e. RGS only applies watermark bias to complete sequences, rather than at every token. For \citet{pmlr-v202-kirchenbauer23a}, who operate without privacy constraints, watermarking at the sequence level is computationally expensive and unnatural. Identifying the set of potential green and red sequences requires a large Monte Carlo sample and provides little benefit, as injecting small distortions at each token generation step already provides exponentially strong detection signal. 

In contrast, for differentially private watermarking, we have the opposite problem. Token-level watermarks lead to hard-to-control privacy leakage, as the information that we leak is a sequence of red-green labels whose length equals the length of the sequence, which is a high-dimensional bit string that is difficult to keep private. This challenge is analogous to the observation in locally differentially private frequency estimation that the optimal estimator is equivalent to a one-bit randomized response-style estimator, as that is the minimal information that needs to be leaked to enable frequency estimation \citep{duchi2013privacy,DBLP:journals/corr/KairouzOV14}. Because of this, sequence-level watermarking is far more natural in our setting, and we encode exactly one red-green bit for each sequence. 

We begin with a naive implementation of this sequence-level approach: given a prompt, we first draw $M$ candidate completions in parallel from the underlying base model, partition them into green and red bins with a keyed hash function, and reweight them appropriately to select the final response. The full pseudocode is presented in Algorithm \ref{alg:rgs-generation}.

\begin{algorithm}
\caption{RGS generation}
\label{alg:rgs-generation}
\begin{algorithmic}[1]
\Require Prompt $x$, candidate budget $M$, privacy budget $\eps$,
         keyed hash $h_k$ with parameter $\pi_0$
\State Draw $\tilde{Y}_1,\ldots,\tilde{Y}_M
       \overset{\mathrm{iid}}{\sim} \LM(\cdot\mid x)$.
\State Deduplicate to $\{(Y_j,c_j)\}_{j=1}^m$;
       set $p_j\gets c_j/M$.
\State $\mathcal{G}\gets\{j:h_k(Y_j)=G\}$;
       $P_G\gets\sum_{j\in\mathcal{G}}p_j$.
\State $Q_G\gets
       \min\{e^\eps P_G,\,
             1-e^{-\eps}(1-P_G)\}$.
\If{$P_G\in\{0,1\}$}
    \State Set $q_j \gets p_j$ for all $j\in[m]$.
\Else
    \State For each $j\in[m]$, set
    \[
    q_j\gets
    \begin{cases}
    p_j Q_G/P_G, & j\in\mathcal{G},\\[2pt]
    p_j (1-Q_G)/(1-P_G), & j\notin\mathcal{G}.
    \end{cases}
    \]
\EndIf
\State Draw $j^*\sim\operatorname{Categorical}(q_1,\ldots,q_m)$.
\State \Return $Y_{j^*}$.
\end{algorithmic}
\end{algorithm}

Comparing Algorithm \ref{alg:rgs-generation} to the earlier token-level red-green watermark in \eqref{eq:kgw-watermark}, we see that one major difference is in the definition of the reweighting term $Q_G/P_G$, which is not a fixed offset determined by $\delta$, but a term that is set to precisely attain our target privacy budget on either the green or red list probability ratio. With this choice of reweighting, we attain our local DP guarantee.

\begin{proposition}[RGS privacy]
\label{prop:rgs-privacy} 
RGS is locally $\eps$-DP with respect to the base model.
\end{proposition}
The proof follows directly from our choice of reweighting coefficients. We defer the proof to Appendix~\ref{proof:rgs-privacy}.

One major drawback of sequence-level watermarks is that the requirement of drawing $M$ parallel samples for one completion is highly wasteful (compared to token-level ones, which are essentially free). However, another difference between standard watermarking and our setting is that the watermarking strength per sequence is \emph{very} low, and thus the difference in distribution before and after watermarking is very small. This enables highly efficient rejection sampling algorithms for our setting, which downweight all red completions by a multiplicative factor of $e^{-\eps}$. At $\eps=1$, the empirical mean number of samples needed by the rejection sampler per watermarked response is around 1.5 on real data, see Section \ref{sec:empirical-rejection-cost}.

\begin{algorithm}[h] 
\caption{RGS via rejection sampling (RGS-reject)}
\label{alg:rgs-rejection}
\begin{algorithmic}[1]
\Require Prompt $x$, candidate budget $M$, privacy budget $\eps$, keyed hash $h_k$ with parameter $\pi_0$
\State Initialize counter $C=0$
\While{$C < M$}
\State Sample a sequence completion $Y \sim \LM(\cdot \mid x)$ independently
\State Increment the counter $C \leftarrow C+1$
\If{$h_k(Y) = G$} 
\State \Return $Y$.
\Else
    \State With probability $e^{-\eps}$, \Return $Y$
\EndIf
\EndWhile
\State \Return $Y$.
\end{algorithmic}
\end{algorithm}

We show formally that our nominal privacy level $\eps$ and the expected number of samples needed by the rejection sampler are directly linked. The proof is given in Appendix~\ref{proof:rgs-reject-expected-sample-count}. The stronger the privacy, the fewer samples needed. 

\begin{fact}[RGS-reject expected sample count]
\label{fact:rgs-reject-sample-count}
The expected number of samples needed for the rejection sampler in Algorithm \ref{alg:rgs-rejection} is at most $1/(\Pr[h_k(Y)=G \mid k] + e^{-\eps} \Pr[h_k(Y)=R \mid k])$.
\end{fact}

As this is a rejection sampler, there is a natural bound on the probability ratio between the original distribution and our sampler, giving a local DP guarantee. The proof of the following privacy guarantee is given in Appendix~\ref{proof:rgs-reject-privacy}.

\begin{proposition}[RGS-reject privacy]
\label{prop:rgs-reject-privacy}
The rejection sampler in Algorithm \ref{alg:rgs-rejection} is $\eps$-DP.
\end{proposition}

We remark that the two samplers use different reweighting rules. The rejection sampler upweights the probability of green candidates relative to red ones by a fixed multiplicative factor of $e^\eps$. The parallel sampler instead uses the empirical green fraction $P_G$ from the $M$ parallel samples to calculate the largest reweighted green probability allowed by the privacy constraints. To see the difference between the weighting rules and to remove the finite budget effect for later analyses, we first state convergence properties of the samplers.

\begin{fact}[Finite-sample convergence]
\label{fact:rgs-convergence}
For any key $k$, denote $g=\Pr[h_k(Y)=G \mid k]$ as the probability of sampling a green sequence from the underlying base model distribution. Let $q_{\mathrm{par}}^M$ and $q_{\mathrm{rej}}^M$ denote the probabilities that the parallel and rejection samplers return a green sequence with candidate budget $M$. Then, with $M\to\infty$,
\[
q_{\mathrm{par}}^M \to \min\{e^\eps g,\,1-e^{-\eps}(1-g)\}, \qquad q_{\mathrm{rej}}^M \to \frac{e^\eps g}{e^{\eps}g + 1-g}.
\]
\end{fact}
The proof is deferred to Appendix \ref{proof:rgs-convergence}. In the infinite budget limit, because $\frac{e^\eps g}{e^{\eps}g + 1-g} \leq  \min\{e^\eps g,\,1-e^{-\eps}(1-g)\}$, the parallel sampler can achieve a larger green probability, while the rejection sampler achieves much better computational efficiency by stopping early.

\subsection{Connection to Randomized Response}
We might view RGS as a type of randomized response (RR) for watermarking, since both methods privatize the true attribute via a randomized one-bit report. While broadly this is correct, there are two major conceptual differences between RR and our watermarking setting that lead to complexities. First, while RR can randomize for both the group with the sensitive attribute and the group without, RGS can only randomize for the watermarked group, since the unwatermarked group is not generated by the watermark decoder. 

Additionally, RGS uses a keyed hash function where a fixed key always assigns the same color to the same sequence, so an analyst can recover the report from the observed text. The construction of the keyed hash function and associated assumptions are discussed in the next section and Appendix \ref{sec:prf-construction}. With the key fixed, the noise in the report now comes from the randomness in sampling the underlying LM. This comes with an important consequence: the extent to which the randomization can be controlled depends on the inherent properties of the underlying base model distribution. The higher the entropy, the better control we have over the randomization, and the more accurate the prevalence estimate can be. We discuss the implications on estimation and design choices in detail in the next section.

\section{Estimation}
\label{sec:estimation}
Having established privacy, we turn to the statistical objective of estimating watermark prevalence. Let $Y_1, \ldots, Y_n$ be a mixture of watermarked and unwatermarked texts, with $W_i \in \{0,1\}$ indicating whether $Y_i$ is watermarked. We define $\lambda = \frac1n \sum_{i=1}^n W_i$ as the true prevalence of watermarked texts in the mixture. Finally, writing $W=(W_1,\ldots,W_n)$, we assume that $Y_1,\ldots,Y_n$ are mutually independent conditional on $W$ and the shared key $k$.

Denote the index sets of watermarked and unwatermarked texts as
$I_1=\{i:W_i=1\}$ and $I_0=\{i:W_i=0\}$. For each $i \in [n]$, define $p_i$ as the reference distribution for the text. For $i \in I_0$, $p_i$ may be any distribution over text sequences and need not come from a language model. For $i \in I_1$, $p_i$ is the \textit{underlying base model distribution} without the watermark decoder $\Gamma$. For example, if the text is generated by prompting the watermarked LM with a prompt $x$, then $p_i$ is the model distribution $p(\cdot|x)$. Throughout, $W$ and the distributions $\{p_i\}_{i=1}^n$ are fixed, and we suppress the conditioning on these quantities.

\subsection{Moment Estimator}

The similarity of RGS to randomized response naturally motivates the use of a moment estimator for prevalence estimation. To construct this
estimator, we first derive the expected fraction of green texts with and without the watermark. 

For each $i \in [n]$ and any fixed key $k$, denote the probability
of sampling a green sequence under the \emph{unwatermarked} distribution $p_i$ as 
\begin{equation}
\label{eq:g_i}
    g_i(k) = \sum_{y \in \gY} p_i(y)\1\{h_k(y)=G\}.
\end{equation} 
For \emph{watermarked} texts $i \in I_1$, the probability of sampling a green sequence is upweighted, which we define as 
\begin{equation}
\label{eq:q_i}
    q_i(k) = \Pr[h_k(\Gamma(k;p_i))=G\mid k].
\end{equation}

Let $H_i=\1\{h_k(Y_i)=G\}$ and
$\bar H=\frac1n\sum_{i=1}^n H_i$. Under the conditional
independence assumption, the indicators $H_i$ conditional on $W$ and $k$ are independent Bernoulli random variables with success probabilities $g_i(k)$ for $i\in I_0$ and $q_i(k)$ for $i\in I_1$ respectively. Therefore,
\begin{align*}
    \E[\bar H\mid k] = \frac{1}{n} \left( \sum_{i\in I_1} q_i(k)  + \sum_{i\in I_0} g_i(k) \right).
\end{align*}

Note that for a fixed $k$, the probabilities $g_i(k)$ and $q_i(k)$ are generally not knowable without assumptions about the underlying distributions $p_i$. To enable estimation, RGS uses a coloring scheme such that each $g_i(k)$ has a known expectation over random keys and then uses $\E_k[g_i(k)]$ as a plug-in for the unknown $g_i(k)$'s.

Ideally, we would pick $k$ uniformly at random, and any sequence would be equally likely to be green or red. However, to enable later detection, we must fix $k$ and have a deterministic coloring as a function of this key. To balance these two competing goals, we implement coloring by thresholding the output of a keyed hash, modeled as a pseudorandom function (PRF) \citep{ggm-prf}. We can then sample the key $k$ once uniformly at random, and reuse it across all texts. The details of the construction can be found in Appendix~\ref{sec:prf-construction}. As a consequence of this construction, we adopt the following assumption for our statistical analysis:
\begin{assumption}[Ideal coloring]
\label{assumption:prf-ideal-coloring}
For every finite collection of distinct sequences
$y_1,\ldots,y_m$ fixed independently of the random key $k$, the
indicators
\[
    \mathbf{1}\{h_k(y_1)=G\},\ldots,
    \mathbf{1}\{h_k(y_m)=G\}
\]
are independent $\operatorname{Bernoulli}(\pi_0)$ random variables.
\end{assumption}

In the formal PRF definition, these are only computationally indistinguishable \citep{ggm-prf}, but we omit the negligible distinguishing advantage from our notation. 

Under this idealization, we can characterize both the key-averaged green probability and its variation across keys. The following lemma also accounts for dependence between $g_i(k)$ and $g_j(k)$ arising from the use of the same key. Its proof is given in Appendix~\ref{proof:lemma:prf-identities}.

\begin{lemma}
\label{lemma:prf-null-green-prob}
Let $p_i$ and $p_j$ be fixed distributions over $\gY$ independent of the key $k$, 
and write
\[
    \|p_i\|_2^2 = \sum_{y \in \mathcal{Y}} p_i(y)^2,
    \qquad
    \langle p_i,p_j\rangle
    = \sum_{y \in \mathcal{Y}} p_i(y)p_j(y).
\]
Under Assumption~\ref{assumption:prf-ideal-coloring},
\begin{align*}
    \mathbb{E}_k[g_i(k)]
        &= \pi_0,\\
    \operatorname{Var}_k[g_i(k)]
        &= \pi_0(1-\pi_0)\|p_i\|_2^2,\\
    \operatorname{Cov}_k[g_i(k),g_j(k)]
        &= \pi_0(1-\pi_0)\langle p_i,p_j\rangle.
\end{align*}
\end{lemma}

Lemma \ref{lemma:prf-null-green-prob} links the second moments of $g_i$ with the diversity of the distributions $p_i$. We can see that $\|p_i\|_2^2$ is the probability that two independent draws from $p_i$ yield the same sequence, which is a measure of (the lack of) entropy of $p_i$, while $\langle p_i,p_j\rangle$ is the probability that independent draws from $p_i$ and $p_j$ produce the same sequence, which measures the (lack of) ``diversity'' of the underlying distributions across individuals.

The variance identity in the lemma suggests that if the underlying distributions $p_i$ have high entropy, and thus low $\|p_i\|_2$, then the green probability $g_i$ would be closer to $\pi_0$ on average. In a non-atomic idealization with $\|p_i\|_2^2 \to 0$, the green probability $g_i$ converges in probability to the null frequency $\pi_0$. Therefore, with $M \to \infty$ in both the parallel and the rejection samplers, the corresponding watermarked green probability is 
\begin{equation}
\label{eq:pi1-both-samplers}
\begin{split}
    \pi_1 = 
    \begin{cases} \min\{e^\eps \pi_0,\, 1-e^{-\eps}(1-\pi_0)\}, & \text{for the parallel sampler},\\[2pt]
    \displaystyle \frac{e^\eps\pi_0}{e^\eps\pi_0+1-\pi_0}, & \text{for the rejection sampler}.
    \end{cases}
\end{split}
\end{equation}
In the ideal case where $g_i = \pi_0$ for every $i$, we have that $\tilde\lambda = \displaystyle \frac{\bar H-\pi_0}{\pi_1-\pi_0}$ is an unbiased estimator of $\lambda$ conditional on $k$ and $W$, exactly mirroring the moment estimator for randomized response. We use the clipped version of this estimator: 
\begin{equation}
\label{eq:rgs-estimator}
\hat \lambda =\mathrm{proj}_{[0,1]} \left( \frac{\bar H-\pi_0}{\pi_1-\pi_0} \right).
\end{equation}

On the other hand, when the underlying distributions do not have perfect entropy, our $g_i(k)$ have nonzero variance over the choice of $k$, and there is no guarantee that the average unwatermarked green frequency will converge to $\pi_0$, resulting in a biased estimator. We now characterize this estimator error that arises from finite watermark entropy and decompose it into bias and variance terms.

\begin{proposition}[Key-conditional MSE]
\label{prop:conditional-mse}
Fix any key $k$. Fix $W = (W_1, \ldots, W_n)$, and suppose that $Y_i \mid W_i$ are conditionally independent. Let
$I_1=\{i:W_i=1\}$, $I_0=\{i:W_i=0 \}$. Assume $0<\lambda<1$, so that both $I_0$ and $I_1$ are nonempty. Define 
\begin{align*}
    V(k) &:= \frac{1}{n}\sum_{i=1}^n
        \bigl(H_i - \E[H_i \mid k]\bigr), \\
    B(k) &:= \frac{1}{|I_1|}\sum_{i\in I_1}
        \bigl(q_i(k) - \pi_1\bigr), \\
    U(k) &:= \frac{1}{|I_0|}\sum_{i\in I_0}
        \bigl(g_i(k) - \pi_0\bigr).
\end{align*}
Then,
\begin{equation}
\label{eq:conditional-mse}
\begin{aligned}
    \E[(\hat\lambda - \lambda)^2 \mid k]
    &\leq \frac{1}{(\pi_1-\pi_0)^2}
       \left[
           \E[V(k)^2 \mid k]
           + \bigl(\lambda B(k) + (1-\lambda)U(k)\bigr)^2
       \right] \\
    &\leq \frac{1}{(\pi_1-\pi_0)^2}
       \left[
           \frac{1}{4n}
           + \bigl(\lambda B(k) + (1-\lambda)U(k)\bigr)^2
       \right].
\end{aligned}
\end{equation}
\end{proposition}

While the variance term $\E[V(k)^2 \mid k]$ vanishes as $O(1/n)$, the bias terms $B(k)$ and $U(k)$ are not guaranteed to vanish, since they depend on the fixed key $k$ and the underlying distributions of the watermarked and unwatermarked texts. We defer the proof to Appendix \ref{proof:prop:conditional-mse}.

\subsection{Null bias and distributional diversity}
The term $U(k)$ measures the bias in the green sequence fraction \emph{marginalized} across all the unwatermarked examples. This bias can be small even when every individual text distribution \(p_i\) has low entropy, provided the distributions are sufficiently diverse across examples.

To see this, observe that Lemma \ref{lemma:prf-null-green-prob} implies
\begin{equation}
\label{eq:Uk-mean-var}
\begin{split}
\E_k[U(k)]  
 & = \frac{1}{|I_0|} \sum_{i\in I_0}(\E_k[g_i(k)]-\pi_0) = 0, \\
\Var_k[U(k)]
 & = \frac{\pi_0(1-\pi_0)}{|I_0|^2} \left(\sum_{i \in I_0} \|p_i\|_2^2 + \sum_{i,j \in I_0, i\neq j} \langle p_i, p_j\rangle\right).
\end{split}
\end{equation}

In this case, we can see that the per-prompt entropies can be low (and thus $\|p_i\|_2^2$ can be high), but the variance is dominated by the cross terms $\langle p_i, p_j\rangle$ for $i \neq j$, which can be small as long as each low-entropy distribution $p_i$ is mutually distinct from another. To illustrate this, consider an example where the distributions $p_i$ all put their mass on a single sequence, but they are all different across $i \in I_0$. Then even with a fixed $k$, because the sequences are all different, the colors would average out. Indeed, in this example $\|p_i\|_2^2 = 1$ for all $i$, but $\langle p_i , p_j \rangle = 0$ for all $i \neq j$.

Note that the variance term can also be written as $\Var_k[U(k)] = \pi_0(1-\pi_0) \|\bar p_0\|_2^2,$
where $\bar p_0(y) := \frac{1}{|I_0|}\sum_{i\in I_0} p_i(y), \; \forall y\in\gY$ is the average text distribution across all unwatermarked texts. When the pool contains many diverse text distributions that spread the mass of $\bar p_0$ over many distinct sequences, the unwatermarked bias term $U(k)$ would concentrate around 0 more tightly.

\subsection{Bias terms for watermarked texts}
Unlike $U(k)$ which can go away under a set of $p_i$'s that are each low-entropy but collectively diverse, the bias term for the watermarked texts $B(k)$ could be persistent even with diverse distributions across prompts. Consider the previous example where the $p_i$'s are all concentrated on a single sequence that differs across $i$'s. For a fixed key, for each $i$ we have $g_i(k)=1$ or $g_i(k)=0$ under both samplers, and the corresponding $q_i(k)$ is either $1$ or $0$, whose average need not equal $\pi_1$.
 
More generally, both samplers bias the model towards producing green sequences by taking the original base model green probability $g_i(k)$ and upweighting it via a concave function. Jensen's inequality then implies a non-positive bias in expectation over $k$. 
\begin{fact}[Key-averaged bias]
\label{fact:rgs-jensen}
Under Assumption~\ref{assumption:prf-ideal-coloring}, both samplers have $\E_k[B(k)]\leq 0$.
\end{fact}

\begin{proof}
Let $f$ denote the mapping from base green probability to output green probability in the limit $M\to\infty$:
\[ 
f(x)= \begin{cases} 
\displaystyle \frac{e^\eps x}{e^\eps x+1-x},
& \text{for the rejection sampler}, \\[6pt]
\min\{e^\eps x, 1-e^{-\eps}(1-x)\} & \text{for the parallel sampler}.
\end{cases} 
\]
Therefore, $\pi_1=f(\pi_0)$. For either sampler, a finite candidate budget $M$ satisfies $q_i(k)\leq f(g_i(k))$. Since $f$ is concave and $\E_k[g_i(k)]=\pi_0$, Jensen's inequality gives
\[
\E_k[q_i(k)] \leq \E_k[f(g_i(k))] \leq f(\E_k[g_i(k)]) =f(\pi_0)=\pi_1.
\]
Averaging over $i\in I_1$ yields $\E_k[B(k)]\leq 0$. For a fixed key, however, the average base green probability may differ from $\pi_0$, and $B(k)$ can have either sign.
\end{proof}

\subsection{Bias control}
\label{sec:bias-control}

Despite persistent bias, there exist design choices that can help us mitigate this bias through a bias-variance tradeoff. We show that adjusting the hash function $h_k$ to have a larger null green fraction $\pi_0$ can reduce the bias at the expense of variance, improving total estimation error. 

Consider the quantity $d = \pi_1-\pi_0$, which is the nominal separation in green probability between the watermarked and unwatermarked text, under the assumption that $g_i = \pi_0$ for every single prompt distribution $p_i$. As previously mentioned, this coincides with the randomized response regime, where $d$ is chosen to be the largest value that respects local $\eps$-DP, which minimizes the variance of the estimator in the absence of bias. In contrast, in our watermark setting, a larger nominal separation $d$ might lead to a larger bias that could offset the benefits from decreasing variance. The following proposition illustrates such a bias-variance tradeoff, in the case with sufficiently many pieces of text (i.e. $|I_0| \to \infty$ and $|I_1| \to \infty$) with sufficient diversity across individuals $i$. 

We remark that in the following theoretical analyses, we study the infinite sample budget limit $M\to\infty$ to isolate the effect of the underlying text distributions on estimation error. For both samplers, by Fact \ref{fact:rgs-convergence} the finite-$M$ green probability $q_i(k)$ converges to $f(g_i(k))$. Empirically, the effect of $M$ plateaus very quickly, see Figure \ref{fig:rejection-m-ablation} and Figure \ref{fig:parallel-m-ablation} in the Appendix.

\begin{proposition}[Key-averaged MSE]
\label{prop:total-mse}
Assume $M \to \infty$ for both samplers, and
\begin{align*}
    \|\bar{p}_0\|_2^2  := \frac{1}{|I_0|^2} \sum_{i,j\in I_0} \langle p_i,p_j\rangle \to 0, \qquad  \|\bar{p}_1\|_2^2 := \frac{1}{|I_1|^2} \sum_{i,j \in I_1} \langle p_i,p_j\rangle \to 0.
\end{align*}
Then
\begin{align*}
    \E[(\tilde \lambda - \lambda)^2 ] \leq \frac{1}{d^2} \E[V(k)^2] + \frac{\lambda^2}{d^2} (\E_k[B(k)])^2 + R,
\end{align*}
where the remainder $R \to 0$ as $\|\bar{p}_0\|_2^2$ and $\|\bar{p}_1\|_2^2 \to 0$.
\end{proposition}

The proof is deferred to Appendix \ref{proof:prop:total-mse}. On a high level, this says that when there is enough diversity across different text distributions, only the sampling noise and the squared watermarked bias contribute to the error. 

As illustrated in Fact \ref{fact:rgs-jensen}, the nominal separation cannot be realized fully when the prompt distributions do not have perfect entropy, inducing a downward bias that need not vanish with sample size. This motivates us to mitigate this bias by reducing the nominal separation, which can be done by increasing $\pi_0$. The following proposition formalizes this monotonicity over sampler-specific ranges.

\begin{proposition}[Bias mitigation by tuning $\pi_0$]
\label{prop:bias-monotonicity}
Consider either sampler in the limit $M\to\infty$. For each $\pi_0\in(0,1)$, let $\pi_1$ be the nominal watermarked green probability as in \eqref{eq:pi1-both-samplers}, and write $d=\pi_1 - \pi_0$. Consider the normalized key-averaged bias
\[ d^{-1} \left| \E_k[B(k)] \right|,\]
where the key expectation is taken under ideal coloring with green probability $\pi_0$. Then:
\begin{enumerate}
    \item For the rejection sampler, $d^{-1} \left| \E_k[B(k)] \right|$ is
    non-increasing on $(0,1)$.
    \item For the parallel sampler, $d^{-1} \left| \E_k[B(k)] \right|$ is non-increasing on $[\sigma(-\eps),1)$, where $\sigmoid(-\eps) :=(1+e^\eps)^{-1}$.
\end{enumerate}
\end{proposition}

The proof is deferred to Appendix~\ref{proof:lemma:rejection-sampler-bias-mitigation}. For the parallel sampler, the nominal separation $d$ is maximized at $\pi_0=\sigma(-\eps)$, while for the rejection sampler it is maximized at $\pi_0=\sigma(-\eps/2)$. This result suggests that increasing $\pi_0$ beyond the RR-optimal values could improve bias at some cost of variance. This tradeoff is particularly relevant at large sample sizes, where sampling noise diminishes but systematic bias can persist. Consistent with this observation, we find empirically that the $\pi_0$ that works best at larger sample sizes is often substantially higher than the value that maximizes separation, see Section \ref{sec:empirical-bias-control}.

\section{Experiments}
We now evaluate the empirical effectiveness of the RGS watermark at estimating the prevalence of watermarked texts in a mixture of watermarked and unwatermarked texts. To assess our estimator in both high-entropy and low-entropy settings, we consider a general language modeling task as well as a much harder coding and essay writing task inspired by real-world instruction-following tasks.

\paragraph{Models} Throughout the experiment, we consider modern models and instruction-tuned models, consisting of Qwen3-1.7B, Qwen3-4B, Qwen3-8B, and Llama-3.1-8B-Instruct. 

\paragraph{Shared watermark parameters and setup} For computational reasons, we use the RGS rejection sampler in Algorithm \ref{alg:rgs-rejection} to generate watermarked completions, with a maximum of $M=20$ candidates, and the keyed hash is implemented as a 128-bit BLAKE2b hash \citep{blake2b} of the generated token sequence. Unless explicitly varied, we use $\pi_0=0.5$ throughout this section. Across RGS estimation experiments, including the ablations (see Appendix \ref{sec:ablations}), we average over 20 randomly sampled keys from the space of $128$-bit keys, and 50 Monte Carlo mixture repetitions per key.

\subsection{Internet text (C4)}
\label{sec:c4}

\begin{figure}[htbp]
    \centering
    \includegraphics[width=1\linewidth]{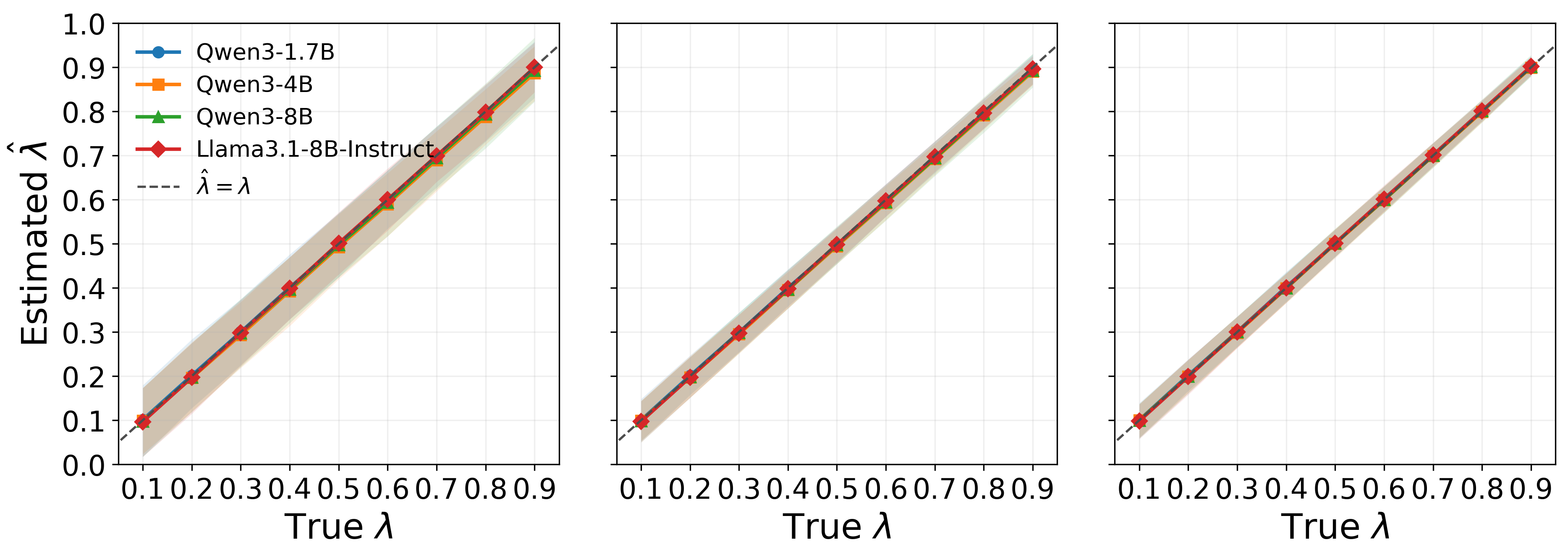}
    \caption{RGS prevalence estimates on C4. Left to right: $\eps=1,2,3$. Markers show means and shading shows 5th--95th percentiles over keys and repetitions.}
    \label{fig:c4}
\end{figure}

We use prompts derived from the Colossal Clean Crawled Corpus (C4), a collection of cleaned web text \citep{c4-paper}. We use the first 500 prompts from the processed C4 dataset in \citet{pan-etal-2024-markllm} as our prompt set. For each prompt, we independently sample 144 base model completions using vLLM with temperature=1 and top-p=1. For each prompt, 64 completions enter the unwatermarked source pool, while the remaining 80 are divided into 4 disjoint blocks of $M=20$ candidates for the rejection sampler. We truncate each candidate to its first $T=50$ generated tokens before hashing and RGS selection, and use the same prefix for detection.

For each $\lambda\in\{0.1,0.2,\ldots,0.9\}$, we construct 50 Monte Carlo mixtures of total size $n=2000$ with $n\lambda$ watermarked and $n(1-\lambda)$ unwatermarked texts sampled uniformly without replacement from the watermarked and unwatermarked source pools. For each mixture, we test RGS generation and estimation with $3$ different privacy budgets $\eps \in \{1,2,3\}$, and repeat the same test using a set of 20 randomly sampled keys.

\paragraph{Accurate frequency estimates with private watermarks} We find that differentially private watermarks provide highly accurate estimates of frequency across a range of $\lambda$, $\eps$, and $n$ values. Figure \ref{fig:c4} shows our estimates of the frequency $\lambda$ as we vary the true frequency $\lambda$ across a wide range. Our estimates show low bias across the entire range, with average absolute biases of $0.0038$, $0.0037$, and $0.001$ for $\eps=1,2,3$ respectively. The 5-95\% percentile bands are generally tight, with average half-widths of $0.072$, $0.040$, and $0.031$ for $\eps=1,2,3$.

These findings hold across a range of $\eps$ values, and our estimator remains accurate even at a fairly stringent $\eps=1$ value. The main challenge of the estimator is maintaining low relative error at small true $\lambda$, but such an estimator may be inherently difficult to construct given the sampling noise inherent to randomized response style frequency estimates.

\paragraph{Sensitivity to watermark and experiment parameters}
To assess how the choices of parameters affect estimator performance, we conduct ablation studies on this C4 experiment. We vary the mixture sample size $n$, privacy budget $\eps$, temperature and generation length, sampler type, candidate budget $M$, and number of distinct prompts in the mixture. The results are presented in Appendix~\ref{sec:ablations}. 

\subsection{Instruction following}
\begin{figure}[h]
    \centering
    \includegraphics[width=0.8\linewidth]{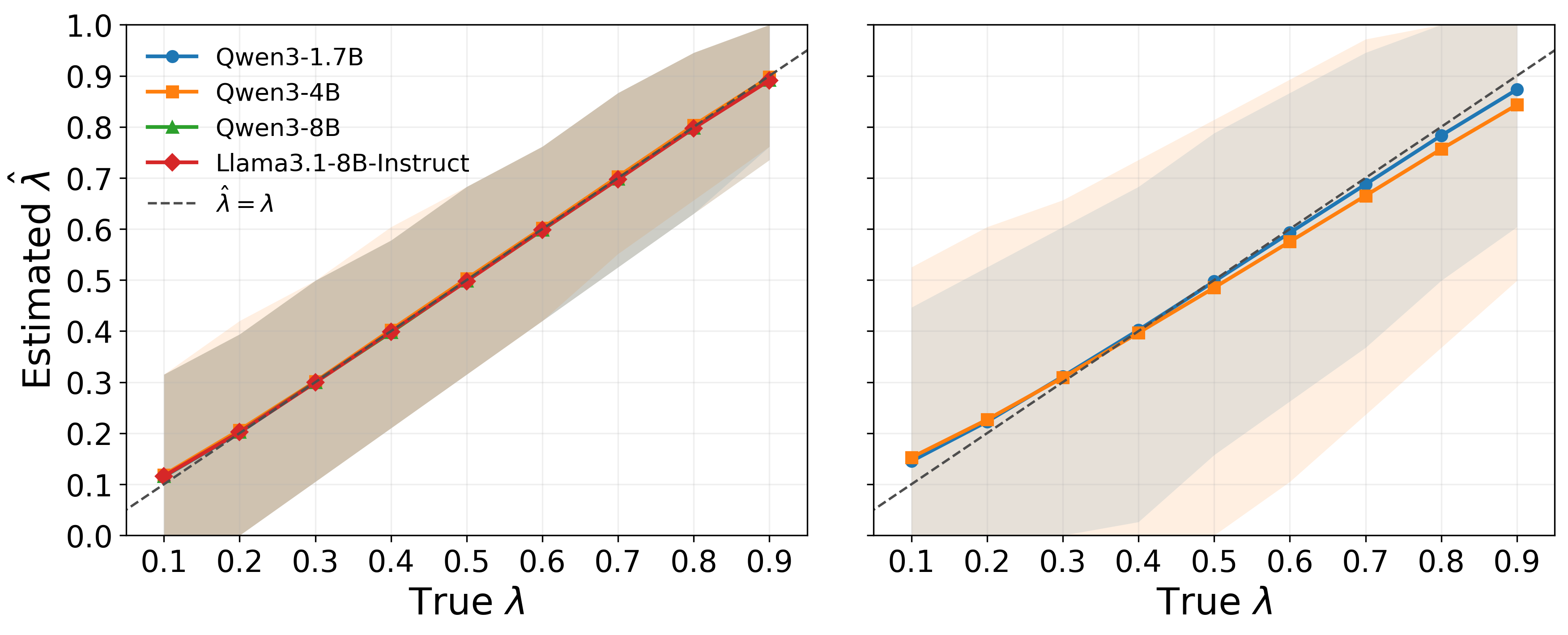}
    \caption{RGS prevalence estimates on OUTFOX (left) and LiveCodeBench (right), with $\eps=2$. Markers show means and shading shows 5th--95th percentiles over keys and repetitions.}
    \label{fig:classroom}
\end{figure}

We then test RGS on a much more challenging setting, such as a small number of students sharing the same homework prompt, where the underlying distributions $p_i$ are low-entropy, and the prompts are highly homogeneous. Throughout this section, we use $\eps=2$. Our setting consists of two datasets: essay writing and coding. 
\paragraph{Essay writing} We randomly select 20 prompts from the OUTFOX dataset consisting of argumentative essay prompts \citep{koike2024outfoxllmgeneratedessaydetection}. Each mixture we test consists of $100\lambda$ texts sampled from the RGS decoder, and $100(1-\lambda)$ from the same base model, all independently sampled under the \textit{same} prompt. 
\paragraph{Coding} We randomly select 20 problems from the LiveCodeBench dataset \citep{jain2024livecodebenchholisticcontaminationfree} and construct the same \textit{one-prompt} mixture as the essay setting. We enable the thinking mode for Qwen3-1.7B and Qwen3-4B and allow them to emit at most 256 thinking tokens before the final answer, which reduces entropy and more accurately reflects realistic LLM use. For watermark generation and detection, we use the first $T=100$ generated tokens, excluding thinking tokens.

\paragraph{Accurate frequency estimates for instruction following settings} We find that even in this extreme stress test where there is only a \emph{single} prompt, and only 100 total samples, we are able to obtain accurate frequency estimates at $\eps=2$ (Figure~\ref{fig:classroom}). The essay setting shows low bias and low overall error across the range of $\lambda$, likely due to the higher per-prompt entropy. In contrast, the coding setting shows higher bias and variance, with visually distinguishable deviation from the unbiased $x=y$ line at low $\lambda$ values. Despite this, the estimates have relatively low bias, and it is promising that even in extreme cases with very low diversity and entropy, we are able to obtain such estimates.

\subsection{Bias control}
\label{sec:empirical-bias-control}
\begin{figure}
    \centering
    \includegraphics[width=1\linewidth]{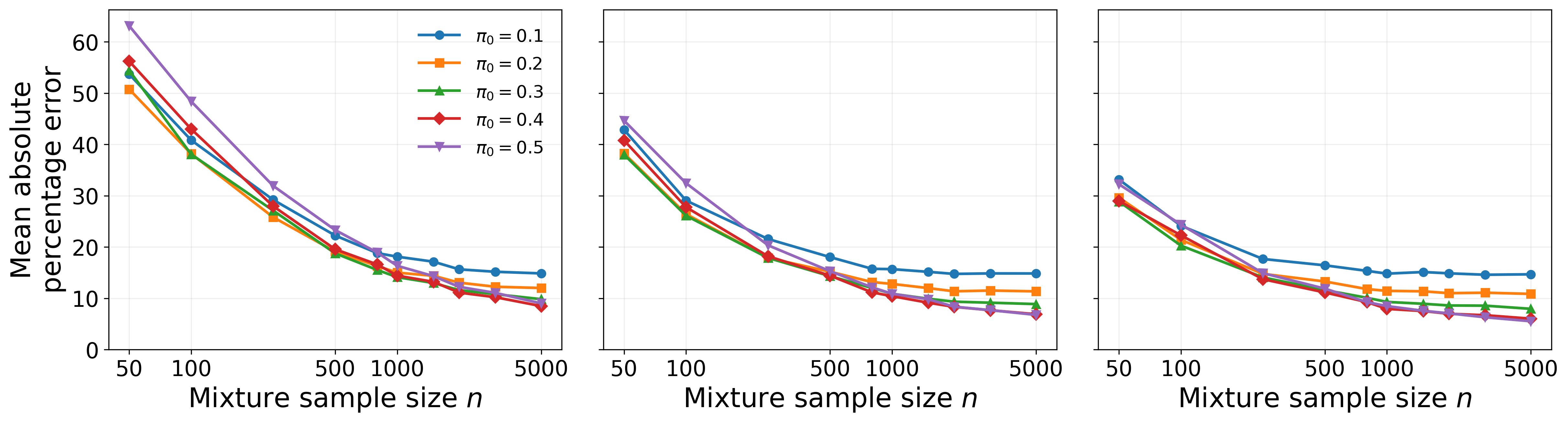}
    \caption{MAPE versus mixture size $n$ for different $\pi_0$. Left to right: $\lambda=0.2,0.3,0.4$.}
    \label{fig:pi0-bias-control}
\end{figure}

We empirically test the effectiveness of the bias-mitigation strategy by tuning $\pi_0$ proposed in Section \ref{sec:bias-control}. We follow the same C4 mixture construction procedure described above, using an expanded set of 1,000 prompts. We sample from Llama-3.1-8B-Instruct with temperature=0.2, and truncate each completion to $T=20$ tokens, reflecting low per-prompt entropy but high prompt diversity. We find that adjusting $\pi_0$ can significantly improve estimation error at large $n$ via the bias variance tradeoff. Figure~\ref{fig:pi0-bias-control} shows the mean absolute percentage error ($100\% \frac{|\hat \lambda -\lambda|}{\lambda}$) averaged over all mixtures and keys, for $\pi_0 \in \{0.1, 0.2, 0.3, 0.4, 0.5\}$ and $\lambda \in \{0.2, 0.3, 0.4\}$, at $\eps=2$. In this setting, the value of $\pi_0$ that optimizes the separation is $\sigmoid(-\eps/2)=0.27$. We observe that at small sample sizes, smaller $\pi_0$ generally achieves lower error, but gradually gets taken over by bigger values of $\pi_0$. In particular, for $\pi_0=0.1$ and $0.2$, the errors plateau quickly across $\lambda$'s. This corroborates our theory that bias persists, and at large sample sizes where bias dominates the total error, a larger $\pi_0$ value achieves a better tradeoff.

\subsection{Computational efficiency}
\label{sec:empirical-rejection-cost}
\begin{figure}
    \centering
    \includegraphics[width=1\linewidth]{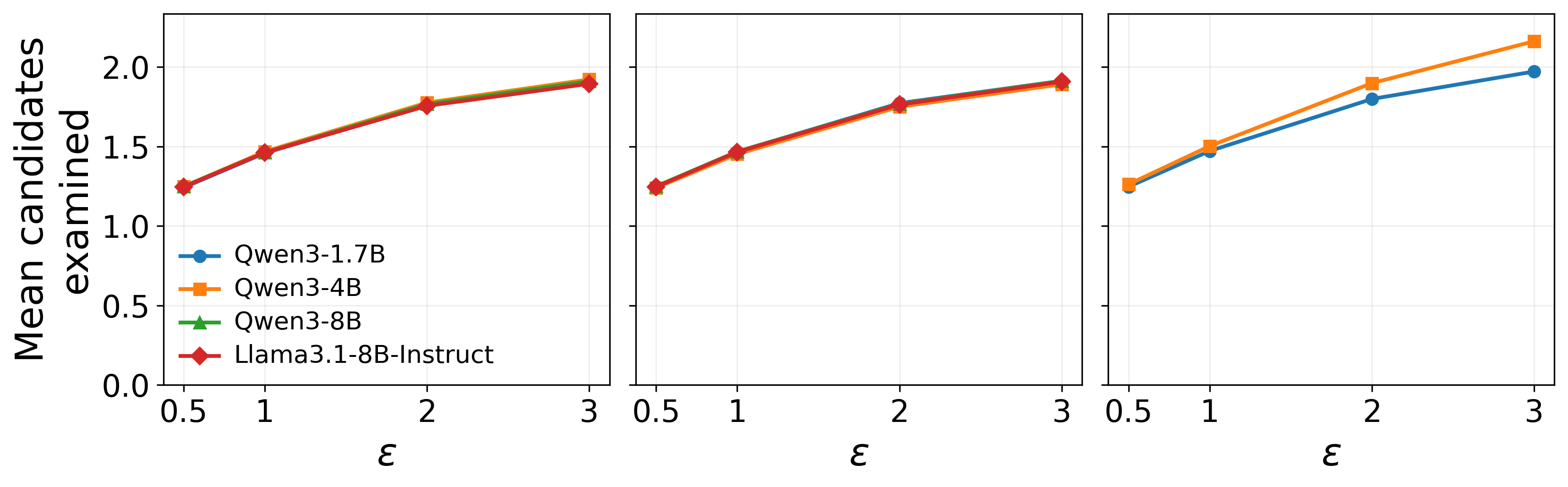}
    \caption{Mean candidates examined by the rejection sampler versus $\eps$. Left to right: C4, OUTFOX, LiveCodeBench.}
    \label{fig:rejection-cost}
\end{figure}

We show empirically that our rejection sampler is very computationally efficient, especially in small $\eps$ regimes, which is consistent with Fact \ref{fact:rgs-reject-sample-count}. In Figure~\ref{fig:rejection-cost}, we plot the mean number of candidates examined until a response is accepted for the rejection sampler, for $\eps\in\{0.5,1,2,3\}$ on C4, OUTFOX, and LiveCodeBench. We use a rejection sampling cap of $M=20$ and $\pi_0=0.5$. Each point averages the cost equally over 20 randomly sampled keys. Across models, we see that only an average of fewer than $1.5$ samples are needed from the underlying base model at small $\eps$ of $0.5$ and $1$, while larger values $\eps=2,3$ require only around 2 samples. 

\section{Discussion}
In this work, we focused on the core challenges behind locally differentially private watermarking, and showed that careful construction of a sequence level, randomized-response style watermark can balance privacy and efficacy. However, there remains important future work in this direction, such as imbuing mild levels of robustness, since currently the sequence-level hashing used by RGS is sensitive to minor edits. Possible mitigations include using locality sensitive hashing \citep{hou-etal-2024-semstamp} or incorporating error correcting codes \citep{christ_pseudorandom_2024}. While a token-level watermark like that of \citet{pmlr-v202-kirchenbauer23a} is more edit-robust and computationally efficient, it remains an important problem to control the total privacy leakage of such token-level watermarks, especially over long, autoregressive sequences. Preliminary investigation by the authors using the optimal DP composition theorem \citep{pmlr-v37-kairouz15} over the text sequence shows some promise for frequency estimation, but the accuracy under a fixed sequence-level privacy budget requires further improvement.

Another direction for future work is mitigating privacy leakage across time from repeated inference, as currently our watermark does not guarantee privacy over unlimited queries by a single user. Although memoized randomized response can protect a sensitive attribute across repeated queries by reusing a previously randomized report \citep{rappor}, they apply randomization to both groups with and without the sensitive attribute. In the watermark setting, since the generation process of the unwatermarked group is outside our control, reusing the same noisy report could introduce a persistent shift in detection signal away from the unwatermarked baseline, revealing the watermark status over time.

\section*{Acknowledgements}
We thank members of the Hashimoto group for useful discussions. ELH was supported by a Stanford Graduate Fellowship. TH was supported by a grant by HAI, gifts from Open Philanthropy, Amazon, Schmidt Sciences, Hudson River Trading, the Tianqiao and Chrissy Chen Foundation and a grant under the NSF CAREER IIS-2338866, ONR N00014-24-1-2609, and DARPA Cooperative Agreement HR00112520013. This work does not necessarily reflect the position or policy of the government and no official endorsement should be inferred.

\bibliographystyle{ims}
\bibliography{bib}

\clearpage
\appendix
\etocdepthtag.toc{appendix}
\etocsettagdepth{main}{none}
\etocsettagdepth{appendix}{subsubsection}
\renewcommand{\contentsname}{Appendix contents}
\etocsetnexttocdepth{subsubsection}
\tableofcontents
\clearpage
\section{Keyed hash construction}
\label{sec:prf-construction}
We construct the coloring scheme described in
Section~\ref{sec:estimation} as follows: first, we use a keyed hash to generate
a pseudorandom value for each sequence, then use a deterministic threshold to map that value to a green or red color. Specifically, we use keyed BLAKE2b \citep{blake2b} with a
128-bit digest. Let
$b_k(y)\in\{0,\ldots,2^{128}-1\}$ denote the integer representation
of the digest of an injectively encoded token sequence $y$.
Here, $b_k$ is the keyed function modeled as a PRF. The coloring
function $h_k$ is its thresholded output, with the threshold
chosen to obtain the nominal green probability $\pi_0$:
\[
    h_k(y)=G
    \quad\Longleftrightarrow\quad
    b_k(y)<T,
    \qquad T=\lfloor \pi_0 2^{128}\rfloor.
\]
If $b_k$ is modeled as a uniformly random function, distinct
sequences receive independent colors, each with green probability
\[
    \widetilde{\pi}_0=\frac{T}{2^{128}},
    \qquad
    0 \leq \pi_0-\widetilde{\pi}_0 < 2^{-128}.
\]
We suppress this small rounding error in the analysis and use $\pi_0$ as the ideal green probability.

\section{Proofs}
\label{sec:appendix-proofs}
\subsection{Sampler proofs}
\label{sec:appendix-sampler-proofs}

\subsubsection{Feasible reweighting under privacy}
\begin{lemma}[Feasible reweighting under privacy]
\label{lemma:feasible-reweighting}
Under Algorithm~\ref{alg:rgs-generation}, with $P_G \in (0,1)$,  a choice of $Q_G$ satisfies
\[ e^{-\eps}p_j\leq q_j\leq e^\eps p_j \qquad\text{for all }j\in[m] \]
if and only if
\[ L_\eps(P_G)\leq Q_G\leq U_\eps(P_G), \]
where
\begin{align*}
L_\eps(P_G)&:=\max\{e^{-\eps}P_G,\;1-e^\eps(1-P_G)\},\\
U_\eps(P_G)&:=\min\{e^\eps P_G,\;1-e^{-\eps}(1-P_G)\}.
\end{align*}
\end{lemma}

\begin{proof}
For $P_G \in (0,1)$, we have that $q_j / p_j = Q_G/P_G$ if $h_k(y_j)=G$, and $q_j/p_j = (1-Q_G)/(1-P_G)$ if $h_k(y_j)=R$.
Therefore, $e^{-\eps}p_j\leq q_j\leq e^\eps p_j$ is equivalent to satisfying both of these inequalities:
\begin{align*}
e^{-\eps}P_G&\leq Q_G\leq e^\eps P_G,\\
1-e^\eps(1-P_G)&\leq Q_G\leq1-e^{-\eps}(1-P_G).
\end{align*}
Their intersection is the stated interval $[L_\eps(P_G),U_\eps(P_G)]$.
\end{proof}

\subsubsection{Proof of Proposition \ref{prop:rgs-privacy}}
\label{proof:rgs-privacy}
\begin{proof}
Fix any key $k$, and let $p(\cdot)$ be the base model distribution given the prompt. The RGS decoder chooses $Q_G=U_\eps(P_G)$, which lies in the
feasible interval in Lemma~\ref{lemma:feasible-reweighting}. Therefore, we have 
\[e^{-\eps}p_j\leq q_j\leq e^\eps p_j \qquad\text{for all }j\in[m].\]
Conditional on the candidate set $Y^M$, for any $y\in\gY$ that appears in the pool, let $c_y$ be the number of appearances of $y$ in $Y^M$. For sequences not in the set, $c_y=0$. Then, we have
\[ e^{-\eps}\frac{c_y}{M} \leq \Pr[\Gamma(k;p)=y\mid k,Y^M] \leq e^\eps\frac{c_y}{M}, \qquad \forall y\in\gY. \]

Taking expectations over $Y^M$ and using
$\E_{Y^M}[c_y/M]=p(y)$ gives the $\eps$-local DP guarantee:
\[
e^{-\eps}p(y)
\leq
\Pr[\Gamma(k;p)=y\mid k]
\leq
e^\eps p(y).
\]
\end{proof}

\subsubsection{Proof of Fact~\ref{fact:rgs-reject-sample-count}}
\label{proof:rgs-reject-expected-sample-count}
\begin{proof}
Fix any key $k$.  In each round of the rejection sampling procedure, the probability of accepting a completion $Y$ is 
\begin{align*}
    a := 1 \cdot \Pr[h_k(Y)=G] + e^{-\eps} \Pr[h_k(Y)=R]. 
\end{align*}
Since the $Y$'s are sampled independently from the base model across all $M$ rounds, the acceptance indicators are independent $\Ber(a)$ random variables. Let $X$ be the number of rounds until acceptance, so $X \sim \text{Geometric}(a)$. The sample count needed by the rejection sampler is thus $T = \min(M, X)$, with $\E[T] \leq \E[X] = \frac{1}{a}$.
\end{proof}

\subsubsection{Proof of Proposition~\ref{prop:rgs-reject-privacy}}
\label{proof:rgs-reject-privacy}
\begin{proof}
Fix any prompt and key, and suppress their dependence in the notation. Let $p$ denote the base model distribution given the prompt. For a candidate $Y\sim p$, let
$\acc$ be the event that it gets accepted by the rejection sampler. Since $\Pr[\acc\mid Y=y]\in\{e^{-\eps},1\}$, we also have
$\Pr[\acc]\in[e^{-\eps},1]$. Thus, by Bayes' rule, the distribution of the accepted candidate satisfies, for $p(y)>0$,
\begin{align*}
q(y)& :=\Pr[Y=y\mid\acc] \\
& = p(y)\frac{\Pr[\acc\mid Y=y]}{\Pr[\acc]} \\
& \in[e^{-\eps}p(y),e^\eps p(y)].
\end{align*}
The same bounds hold when $p(y)=0$, since $q(y)=0$ in that case.

The finite candidate budget $M$ only moves the watermarked output distribution toward the base $p$. If round $M$ is reached, its fresh candidate has distribution identical to the base $p$ and is returned by the rejection sampler regardless of color. Writing $r$ as the probability of reaching the final round, the watermarked distribution is therefore
\[
q_M(y)=(1-r)q(y)+rp(y),\qquad r\in[0,1].
\]
This mixture only moves each likelihood ratio toward $1$, so
\[
e^{-\eps}p(y)\leq q_M(y)\leq e^\eps p(y)
\qquad \forall y \in \gY.
\]
\end{proof}

\subsubsection{Proof of Fact~\ref{fact:rgs-convergence}}
\begin{proof}
\label{proof:rgs-convergence}
For the parallel sampler, the empirical green fraction $P_G$ in Algorithm \ref{alg:rgs-generation} converges almost surely
to $g$ by the strong law of large numbers. Conditional on the
candidates and $P_G$, the output is green with probability
$f(P_G)=\min\{e^\eps P_G,1-e^{-\eps}(1-P_G)\}$. Since $f$ is continuous and bounded, $q_{\mathrm{par}}^M=\E[f(P_G)] \to f(g)$ by the bounded convergence theorem.

For the rejection sampler, the probability of accepting a completion on each attempt before the final one is $a=g+e^{-\eps}(1-g)$. Since each attempt is independent, the cumulative green probability is
\[
q_{\mathrm{rej}}^M  = \sum_{t=1}^{M}(1-a)^{t-1}g \to \frac{g}{a} =\frac{e^\eps g}{e^{\eps}g + 1-g}. 
\]
\end{proof}

\subsection{Estimation proofs}
\label{sec:appendix-estimation-proofs}

\subsubsection{Proof of Lemma~\ref{lemma:prf-null-green-prob}}
\label{proof:lemma:prf-identities}
\begin{proof}
Use the shorthand $Z_y=\1\{h_k(y)=G\}$. Under Assumption \ref{assumption:prf-ideal-coloring}, the $Z_y$'s are independent $\Ber(\pi_0)$ random variables. Since
$g_i(k)=\sum_{y\in\gY}p_i(y)Z_y$, we have
\[
\E_k[g_i(k)]
=\sum_{y\in\gY}p_i(y)\E_k[Z_y]
=\pi_0.
\]
Furthermore, independence gives
\begin{align*}
\Cov_k[g_i(k),g_j(k)]
&=\sum_{y,y'\in\gY}p_i(y)p_j(y')
  \Cov_k[Z_y,Z_{y'}]\\
&=\pi_0(1-\pi_0)\sum_{y\in\gY}p_i(y)p_j(y)\\
&=\pi_0(1-\pi_0)\langle p_i,p_j\rangle.
\end{align*}
\end{proof}

\subsubsection{Proof of Proposition~\ref{prop:conditional-mse}}
\label{proof:prop:conditional-mse}
\begin{proof}
With the unclipped moment estimator $\tilde \lambda = \frac{ \bar{H}- \pi_0}{ \pi_1 - \pi_0}$, we have the decomposition
\begin{align*}
    \tilde \lambda - \lambda= \frac{1}{\pi_1-\pi_0} \left[\frac1n \sum_{i=1}^n (H_i - \E[ H_i \mid k]) + \frac1n\sum_{i\in I_1}(q_i-\pi_1) + \frac1n \sum_{i\in I_0}(g_i-\pi_0) \right].
\end{align*}
Denote $V(k) := \frac1n \sum_{i=1}^n (H_i - \E[ H_i \mid k])$.
Then  
\[  \tilde \lambda - \lambda = \frac{1}{\pi_1-\pi_0} 
\bigg[V(k) + \lambda B(k) + (1-\lambda) U(k) \bigg]
\]
Note that $B(k)$ and $U(k)$ are fixed given $k$. The only random quantity is $V(k)$, the sampling noise, with
\begin{align*}
    \E[V(k) \mid k] = 0.
\end{align*}
Then, the conditional MSE is
\begin{align*}
    \E[(\tilde \lambda - \lambda)^2 \mid k] 
    & = \frac{1}{(\pi_1-\pi_0)^2} 
    \left[\E[V(k)^2 \mid k] + 
    \left(\lambda B(k) + (1-\lambda) U(k) \right)^2 \right]. 
\end{align*}
By conditional independence, we have
\begin{align*}
    \E[V(k)^2 \mid k] & = \Var[V(k)\mid k]\\ 
    & = \Var\left[\frac1n \sum_{i=1}^n H_i \mid k\right] \\
    & = \frac{1}{n^2} \sum_{i=1}^n \Var[H_i \mid k] \\
    & = \frac{1}{n^2} \left( \sum_{i\in I_1} q_i(1-q_i) + \sum_{i\in I_0} g_i(1-g_i)  \right) \\
    & \leq \frac{1}{4n}.
\end{align*}

Since projection onto $[0,1]$ can only decrease the MSE, combining the above we have
\begin{align*}
    \E[(\hat \lambda - \lambda)^2 \mid k] 
    & \leq \frac{1}{(\pi_1-\pi_0)^2} 
    \left[\frac{1}{4n} + 
    \left(\lambda B(k) + (1-\lambda) U(k) \right)^2 \right].
\end{align*}
\end{proof}

\subsubsection{Lemma: Variance of \texorpdfstring{$B(k)$}{B(k)}}
\begin{lemma}[Variance of $B(k)$]
\label{lemma:var-Bk}
Consider both samplers in the limit $M\to\infty$. Let $m=|I_1|>0$. Under Assumption~\ref{assumption:prf-ideal-coloring}, we have
\[
\Var_k[B(k)] \leq \frac{e^{2\eps}\pi_0(1-\pi_0)}{m^2}
\left( \sum_{i\in I_1}\|p_i\|_2^2
    + \sum_{\substack{i,j\in I_1\\ i\ne j}}\langle p_i,p_j\rangle
\right).
\]
\end{lemma}

\begin{proof}
\label{proof:lemma:var-Bk}
For either sampler, write $q_i(k)=f(g_i(k))$, where
\[
f(x)=
\begin{cases}
\displaystyle \frac{e^\eps x}{e^\eps x + 1-x},
    & \text{for the rejection sampler},\\[6pt]
\min\{e^\eps x,\,1-e^{-\eps}(1-x)\},
    & \text{for the parallel sampler}.
\end{cases}
\]
Both maps are $e^\eps$-Lipschitz on $[0,1]$.
For the rejection sampler, this follows from
\[
0\le f'(x)=\frac{e^\eps}{(1+(e^\eps-1)x)^2}\le e^\eps.
\]
For the parallel sampler, $f$ is continuous and piecewise linear
with slopes $e^\eps$ and $e^{-\eps}$.

Write $Z_y=\1\{h_k(y)=G\}$, so that $g_i(k)=\sum_y p_i(y)Z_y$, and $B(k)$ is a function of $Z_y$:
\[B(k) = \frac1m \sum_{i\in I_1} f\left( \sum_{y \in \gY} p_i(y) Z_y \right) - f(\pi_0).\]
Under Assumption~\ref{assumption:prf-ideal-coloring}, the $Z_y$ are
independent $\Ber(\pi_0)$ variables.  Suppose $B(k)'$ is the variable obtained by replacing $Z_y$ by an i.i.d. copy $Z_y'$. By the Lipschitz bounds, this changes $B(k)$ by at most $c_y := \frac{e^\eps}{m}\sum_{i\in I_1}p_i(y)$, i.e. $|B(k) - B(k)'| \leq c_y$.

Note that this change is only nonzero
if $Z_y\neq Z'_y$, and since they are independent $\Ber(\pi_0)$ variables, this event has probability $2\pi_0(1-\pi_0)$. Therefore, by the Efron-Stein inequality \citep{Boucheron2004}, we have 
\begin{align*}
\Var_k[B(k)] & \leq \frac{1}{2} \sum_{y \in \gY} \E_k[|B(k) - B(k)'|^2] \\
& \leq \tfrac12\sum_{y} 2\pi_0(1-\pi_0)\,c_y^2 \\
& = \frac{e^{2\eps}\pi_0(1-\pi_0)}{m^2}\sum_{i,j\in I_1}\langle p_i,p_j\rangle.
\end{align*}
\end{proof}

\subsubsection{Proof of Proposition~\ref{prop:total-mse}}
\begin{proof}
\label{proof:prop:total-mse}
Taking expectation over the key $k$ in \eqref{eq:conditional-mse}, we can decompose the error to
\begin{equation}
\label{eq:total-mse}
\begin{split}
    & \E[(\tilde \lambda - \lambda)^2]\\
    = &  \frac{1}{d^2}\E[V(k)^2] 
    + \frac{\lambda^2}{d^2} \E_k[B(k)^2] + 
    \frac{(1-\lambda)^2}{d^2} \E_k[U(k)^2] + 
    \frac{2\lambda(1-\lambda)}{d^2} \E_k[B(k)U(k)].
\end{split}
\end{equation}
We bound the last three terms of the equation above. Since $\E_k[U(k)]=0$, we have
\begin{align*}
     \E_k[U(k)^2] & = \Var_k[U(k)] = \pi_0(1-\pi_0) \|\bar{p}_0\|_2^2.
\end{align*}
Then, we bound the cross term using Cauchy-Schwarz:
\begin{align*}
\E_k[B(k)U(k)] 
     & \leq |\E_k[B(k)U(k)] |\\
     & \leq \sqrt{\E_k[B(k)^2]} \sqrt{\E_k[U(k)^2]} \\
     & = \sqrt{\E_k[B(k)^2]} \sqrt{\pi_0(1-\pi_0) \|\bar{p}_0\|_2^2} \\
     & \leq \sqrt{\pi_0(1-\pi_0)}\, \|\bar{p}_0\|_2,
\end{align*}
Finally, we bound $\Var_k[B(k)]$ using Lemma \ref{lemma:var-Bk}:
\begin{align*}
\E_k[B(k)^2] & = \Var_k[B(k)] + (\E_k[B(k)])^2 \\
     & \leq e^{2\eps}\pi_0(1-\pi_0) \|\bar{p}_1\|_2^2 + (\E_k[B(k)])^2.
\end{align*}
Substituting the above into \eqref{eq:total-mse} gives the desired
inequality with remainder $R$. Every term of $R$ is proportional to
$\|\bar p_0\|_2$, $\|\bar p_0\|_2^2$ or $\|\bar p_1\|_2^2$, so $R \to 0$
under the assumption.
\end{proof}

\subsubsection{Proof of Proposition~\ref{prop:bias-monotonicity}}
\begin{proof}
\label{proof:lemma:rejection-sampler-bias-mitigation}
Write $m = |I_1|$ and without loss of generality, suppose $m>0$. We consider both samplers.
\textbf{Rejection Sampler.}
Consider the rejection sampler in the limit $M\to\infty$, so that
$q_i(k)=f(g_i(k))$, where
\[
f(x) = \displaystyle \frac{e^\eps x}{e^\eps x + 1-x}.
\]
Then, for any $x, y \in [0,1]$, we have
\begin{align*}
    f(x) - f(y) = \frac{e^\eps(x-y)}{(1+(e^\eps-1)x)(1+(e^\eps-1)y)}. 
\end{align*}
We use the shorthand $Z_y = \1\{h_k(y)=G\}$, which implies $g_i = \sum_{y \in \gY} p_i(y) Z_y.$ Then, the unnormalized bias is 
\begin{align*}
    \E_k\left[ f(g_i) - f(\pi_0)     
    \right] 
    & = \E_k\left[  \frac{e^\eps(g_i - \pi_0)}{(1+(e^\eps-1) g_i)(1+(e^\eps-1) \pi_0)}   
    \right] \\
    & = \frac{e^\eps}{1+(e^\eps-1)\pi_0} \sum_{y \in \gY} p_i(y) \E \left[ \frac{ Z_y - \pi_0}{1+(e^\eps-1) g_i} \right].
\end{align*}
Since both the numerator and the denominator in the integrand depend on $Z_y$, which is a Bernoulli random variable with mean $\pi_0$, we use the leave-one-out trick to tease out the dependence of the whole expectation on $\pi_0$.  

Denote $g_i^{-y} = \sum_{y' \neq y}p_i(y') Z_{y'}$. Then, $g_i = g_i^{-y} + p_i(y)Z_y$. Note that conditional on $g_i^{-y}$, $Z_y \sim \Ber(\pi_0)$. Then, 
\begin{align*}
    \E_k \left[ \frac{ Z_y - \pi_0}{1+(e^\eps-1) g_i} \right] & = 
    \E \left[ \E \left[ \frac{ Z_y - \pi_0}{1+(e^\eps-1) (g_i^{-y} + p_i(y)Z_y)}  \mid g_i^{-y} \right]  \right] \\
    & = \E\left[ 
     \pi_0 \frac{1- \pi_0}{1+(e^\eps-1) (g_i^{-y} + p_i(y))} + (1-\pi_0) \frac{-\pi_0}{1+(e^\eps-1)g_i^{-y}}
    \right] \\
    & =  -\pi_0(1-\pi_0)(e^\eps-1)p_i(y) \E_k\left[\phi(g_i^{-y}) \right],
\end{align*}
where 
\[\phi(g_i^{-y}) = \frac{1}{(1+(e^\eps-1) (g_i^{-y} + p_i(y)))(1+(e^\eps-1)g_i^{-y})}. \] 
Consolidating, the magnitude of the normalized bias is 
\begin{align*}
    \frac{\E_k[-B(k)]}{f(\pi_0) - \pi_0} 
    & = -\frac{1}{f(\pi_0) - \pi_0} \frac1m  \sum_{i\in I_1} \E_k [ f(g_i) - f(\pi_0) ] \\ 
    & = \frac{\pi_0(1-\pi_0)e^\eps(e^\eps-1)}{m(f(\pi_0) - \pi_0)(1+(e^\eps-1)\pi_0)} \sum_{i\in I_1} \sum_{y\in \gY}p_i(y)^2 \E_k \left[\phi(g_i^{-y}) \right] \\
    & = \frac{e^\eps}{m}\sum_{i\in I_1} \sum_{y\in \gY}p_i(y)^2 \E_k \left[\phi(g_i^{-y}) \right].
\end{align*}
Note that $\phi$ is nonincreasing. Also $g_i^{-y} = \sum_{y' \neq y}p_i(y') Z_{y'}$, where each $Z_{y'} \overset{i.i.d.}{\sim}\Ber(\pi_0)$. Since a $\Ber(\pi_0')$ variable stochastically dominates a $\Ber(\pi_0)$ variable for $\pi_0' \geq \pi_0$, by Corollaries 4.2.7 and 4.2.6 of \cite{roch_mdp_2024} we have that $\E_{k}[\phi(g_i^{-y})]$ is nonincreasing in $\pi_0$. Overall, since the multipliers $e^\eps/m$ and $p_i(y)^2$ are nonnegative, the magnitude of the normalized bias $\frac{\E_k[-B(k)]}{f(\pi_0) - \pi_0}$ is nonincreasing in $\pi_0$.

\textbf{Parallel Sampler.}
Consider the parallel sampler in the limit $M\to\infty$, so that
$q_i(k)=f(g_i(k))$, where
\[
f(x)=\min\{e^\eps x,\;1-e^{-\eps}(1-x) \}.
\]
Similar to the rejection sampler proof, we use the shorthand $Z_y=\1\{h_k(y)=G\}$, which implies $g_i=\sum_{y\in\gY}p_i(y)Z_y$. Under the ideal coloring assumption, the $Z_y$ are independent $\Ber(\pi_0)$ variables, with $\E_k[g_i]=\pi_0$.

Let $\tau=(1+e^\eps)^{-1}$. Note that $f$ is a piecewise linear function, which can be rewritten as
\begin{equation} 
\label{eq:piecewise-identity}
\begin{split}
f(x) = 1-e^{-\eps}(1-x) -(e^\eps-e^{-\eps})(\tau-x)_+ 
\end{split}
\end{equation}

We consider $\pi_0\in[\tau,1)$. Using the identity in \eqref{eq:piecewise-identity}, we can write the unnormalized bias as
\begin{align*}
\E_k[f(g_i)-f(\pi_0)]
&= \E_k\left[ e^{-\eps}(g_i-\pi_0) -(e^\eps-e^{-\eps})(\tau-g_i)_+ \right]\\
&=-(e^\eps-e^{-\eps})\E_k[(\tau-g_i)_+].
\end{align*}

To isolate the dependence on $\pi_0$, define the nonincreasing function
\[
\psi(x)= \begin{cases}
\displaystyle \frac{\tau-x}{1-x}, & 0\leq x<\tau, \\[6pt]
0, & \tau\leq x\leq1.
\end{cases}
\]
Then $(\tau-g_i)_+=(1-g_i)\psi(g_i)$, so
\begin{align*}
\E_k[(\tau-g_i)_+]
& = \E_k[ (1-g_i) \psi(g_i)]\\
& = \sum_{y\in \gY} p_i(y) \E_k[(1-Z_y) \psi(g_i)].
\end{align*}
The interchange of expectation and summation is valid because
all summands are nonnegative.

Since both factors in the integrand depend on $Z_y$, we use
the leave-one-out trick to isolate this dependence.
Denote $g_i^{-y}=\sum_{y'\ne y}p_i(y')Z_{y'}$.
Then $g_i=g_i^{-y}+p_i(y)Z_y$, and conditional on $g_i^{-y}$,
$Z_y\sim\Ber(\pi_0)$. Therefore,
\begin{align*}
\E_k[(1-Z_y) \psi(g_i)]
& = \E_k \left[ \E_k \left[ (1-Z_y) \psi(g_i^{-y} + p_i(y)Z_y) \mid g_i^{-y}
\right]\right]\\
& = (1- \pi_0) \E_k[ \psi(g_i^{-y})].
\end{align*}
Consolidating, the magnitude of the normalized bias is
\begin{align*}
\frac{\E_k[-B(k)]}{f(\pi_0)-\pi_0}
&=-\frac{1}{m(f(\pi_0)-\pi_0)}
  \sum_{i\in I_1}\E_k[f(g_i)-f(\pi_0)]\\
&=\frac{(e^\eps-e^{-\eps})(1-\pi_0)}
        {m(f(\pi_0)-\pi_0)}
  \sum_{i\in I_1}\sum_{y\in\gY}
  p_i(y)\E_k[\psi(g_i^{-y})]\\
&=\frac{1+e^\eps}{m}
  \sum_{i\in I_1}\sum_{y\in\gY}
  p_i(y)\E_k[\psi(g_i^{-y})],
\end{align*}
where the last equality comes from the fact that $f(\pi_0)-\pi_0=(1-e^{-\eps})(1-\pi_0)$.

Note that $\psi$ is nonincreasing. By the same argument as the rejection sampler case, we have that $\E_{k}[\psi(g_i^{-y})]$ is nonincreasing in $\pi_0$ and thus $\frac{\E_k[-B(k)]}{f(\pi_0)-\pi_0}$ is nonincreasing in $\pi_0$.

\end{proof}

\newpage

\section{Ablations}
\label{sec:ablations}
\suppressfloats[t]
We conduct ablations on the C4 task described in Section~\ref{sec:c4}. Unless varied, we use rejection sampling with $\eps=2$, $n=2000$, $\pi_0=0.5$, a candidate cap of $M=20$, and temperature 1. Error plots report mean absolute percentage error (MAPE), obtained by averaging $100|\hat\lambda-\lambda|/\lambda$ over repeated estimates with 20 randomly sampled keys, where $\hat\lambda$ is the clipped estimator in \eqref{eq:rgs-estimator}.

\subsection{Effect of \texorpdfstring{$\eps$}{epsilon} and \texorpdfstring{$n$}{n}}
We compare the performance of our watermark estimator on C4 data with the clipped RR moment estimator (\eqref{eq:rr-mom-estimator} with clipping onto $[0,1]$) under an ideal randomized response experiment described as follows. 
For each mixture sample size $n$ and prevalence $\lambda$, we fix $m=n\lambda$ watermarked examples and draw independent noisy reports $R_1\sim\operatorname{Binomial}(m,\sigma(\eps))$ and $R_0\sim\operatorname{Binomial}(n-m,0.5)$. We compute $R=R_1+R_0$, estimate prevalence using $\hat\lambda_{\mathrm{RR}}=\mathrm{proj}_{[0,1]}\bigl((R/n-0.5)/(\sigma(\eps)-0.5)\bigr)$, and compute MAPE over 1,000 simulations per condition. This RR benchmark matches the nominal green probabilities of RGS rejection sampling at $\pi_0=0.5$ in the limit of high entropy and infinite candidate budget. 

\begin{figure}[htbp]
    \centering
    \includegraphics[width=1\linewidth]{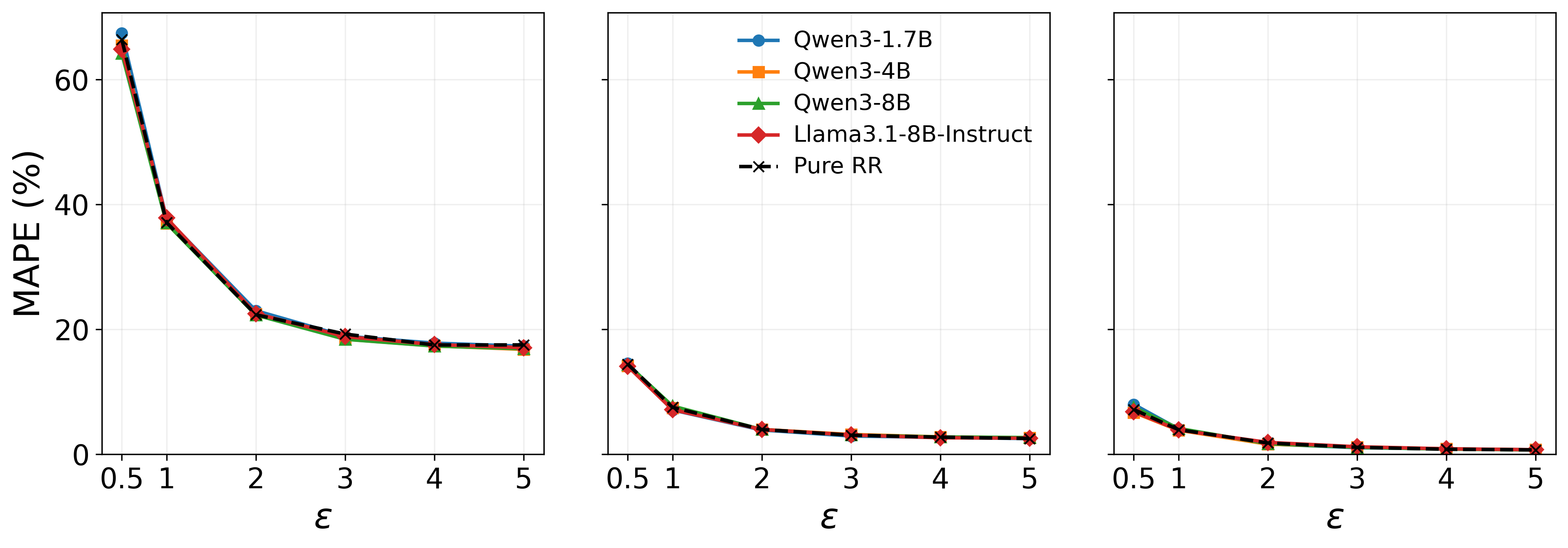}
    \caption{MAPE versus privacy budget $\eps$, with the ideal RR benchmark dashed. Left to right: $\lambda=0.1,0.5,0.9$.}
    \label{fig:epsilon-ablation}
\end{figure}

\begin{figure}[htbp]
    \centering
    \includegraphics[width=1\linewidth]{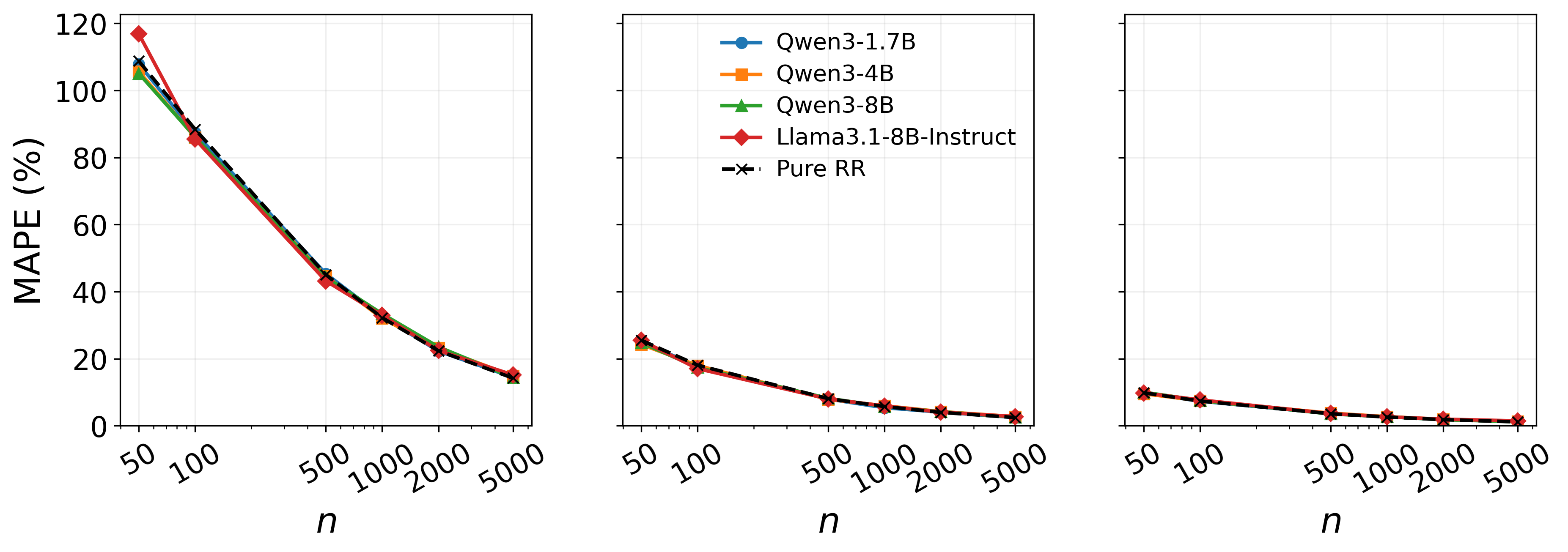}
    \caption{MAPE versus mixture size $n$, with the ideal RR benchmark dashed. Left to right: $\lambda=0.1,0.5,0.9$.}
    \label{fig:n-ablation}
\end{figure}

Figure~\ref{fig:epsilon-ablation} plots MAPE as a function of $\eps$ at a fixed $n=2000$. We observe that the errors decrease with increasing $\eps$, consistent with the looser privacy requirements. Figure~\ref{fig:n-ablation} plots MAPE as a function of sample size $n$ at a fixed $\eps=2$, which shows strong scaling. Across both figures we see that in this high-entropy C4 task, the RGS watermark behaves very similarly to the moment estimator in an ideal randomized response mechanism.

\subsection{Effect of temperature and generation length}
\begin{figure}[htbp]
    \centering
    \includegraphics[width=1\linewidth]{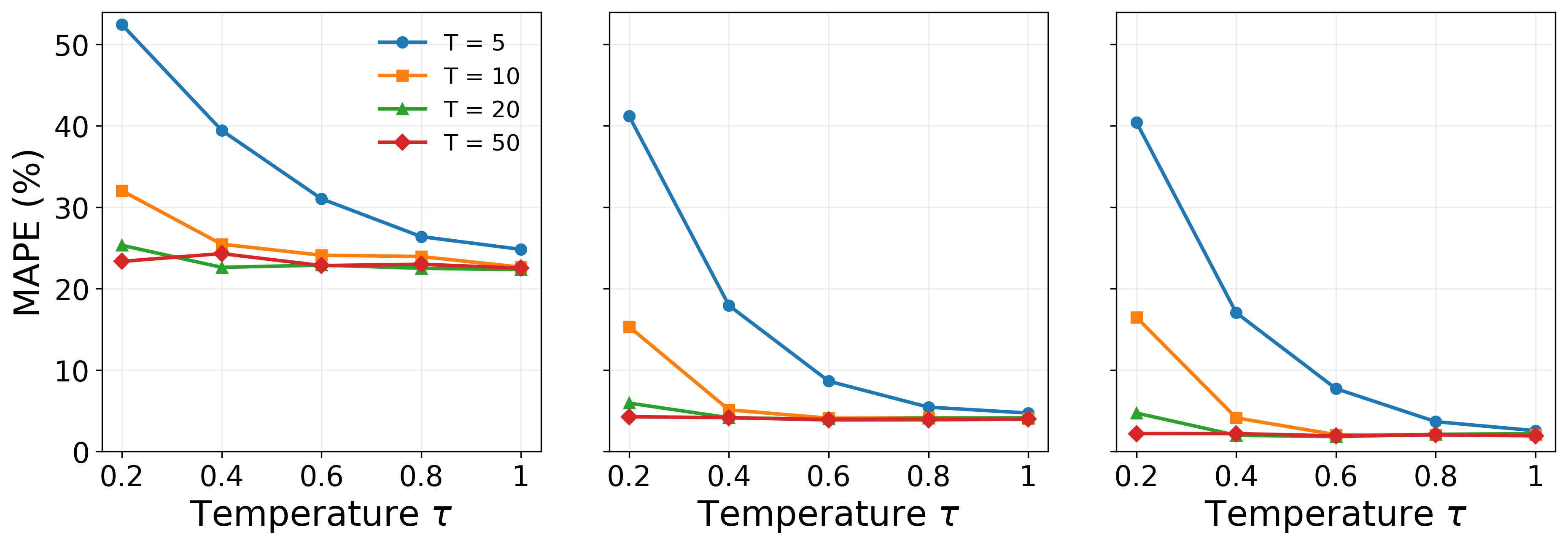}
    \caption{MAPE versus temperature for Llama-3.1-8B-Instruct with prefix lengths $T\in\{5,10,20,50\}$. Left to right: $\lambda=0.1,0.5,0.9$.}
    \label{fig:temp-prefix-ablation}
\end{figure}

Higher temperature and longer generation lengths both increase the entropy of the text distribution. Figure~\ref{fig:temp-prefix-ablation} varies both temperature and the number of generated prefix tokens $T$ used for hashing and selection for Llama-3.1-8B-Instruct. Increasing temperature substantially reduces error for short prefixes, while longer prefixes attain lower error even at low temperatures. These trends agree with the theoretical intuition that higher entropy leads to lower error.

\begin{figure}[htbp]
    \centering
    \includegraphics[width=1\linewidth]{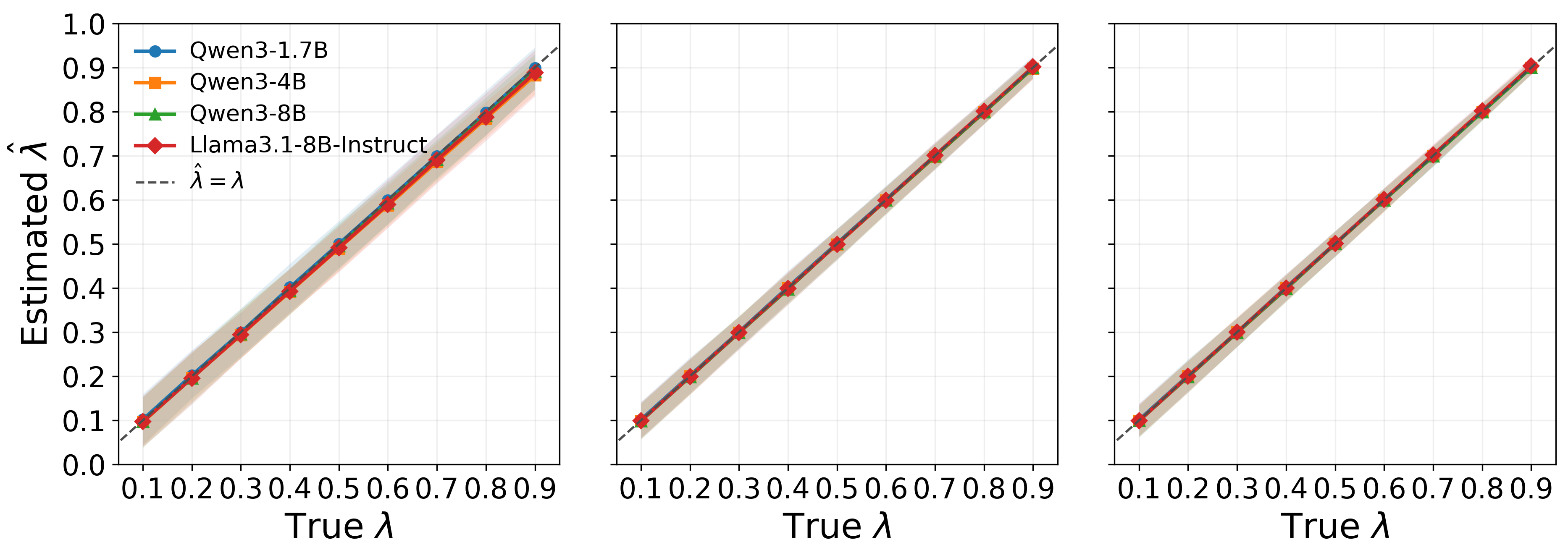}
    \caption{RGS parallel-sampling prevalence estimates on C4. Left to right: $\eps=1,2,3$. Markers show means and shading shows 5th--95th percentiles over keys and repetitions.}
    \label{fig:parallel-ablation}
\end{figure}

\begin{samepage}
\subsection{Parallel versus rejection sampling}
Figure~\ref{fig:parallel-ablation} evaluates the parallel sampler in Algorithm~\ref{alg:rgs-generation} using the same C4 setup as Figure~\ref{fig:c4}. Mean estimates remain close to the true prevalence across all four models, and the percentile bands narrow as $\eps$ increases. Compared with the rejection sampler, the parallel sampler has narrower bands, most visible at $\eps=1$. This is consistent with its larger nominal green probability as described in Section \ref{sec:algorithm}.
\par\end{samepage}
\subsection{Effect of candidate budget}
\begin{figure}[p]
    \centering
    \includegraphics[width=1\linewidth]{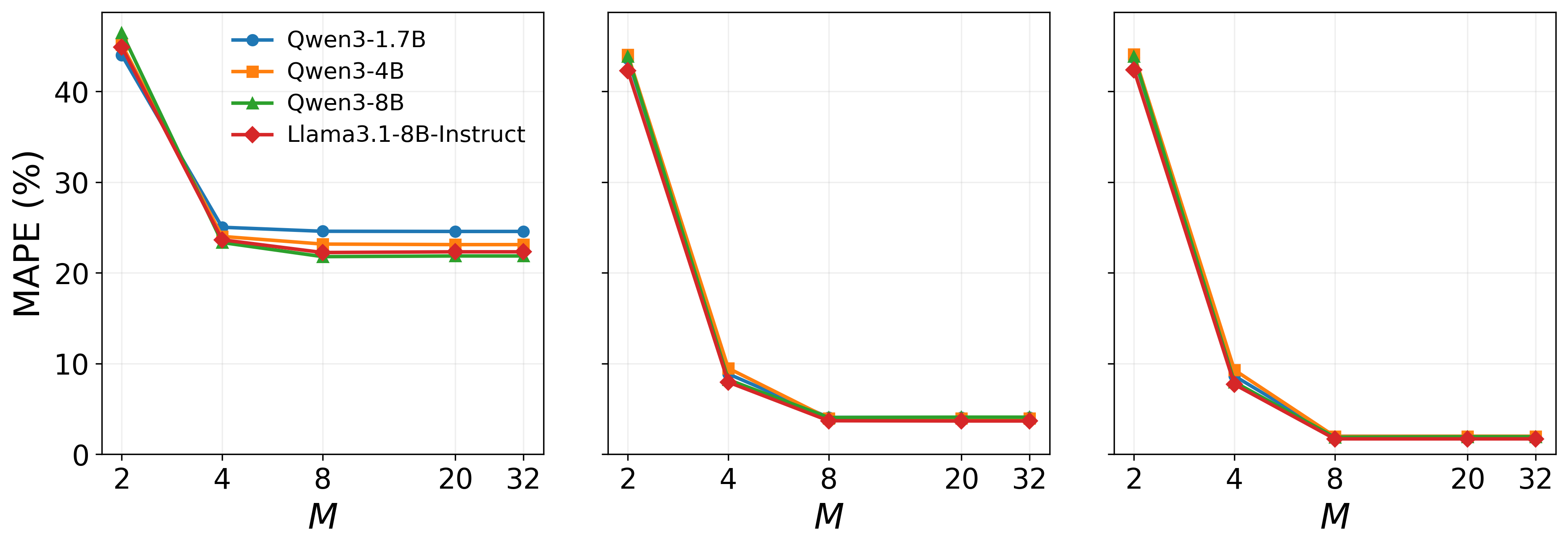}
    \caption{MAPE versus candidate budget $M$ for rejection sampling. Left to right: $\lambda=0.1,0.5,0.9$.}
    \label{fig:rejection-m-ablation}
\end{figure}

\begin{figure}[p]
    \centering
    \includegraphics[width=1\linewidth]{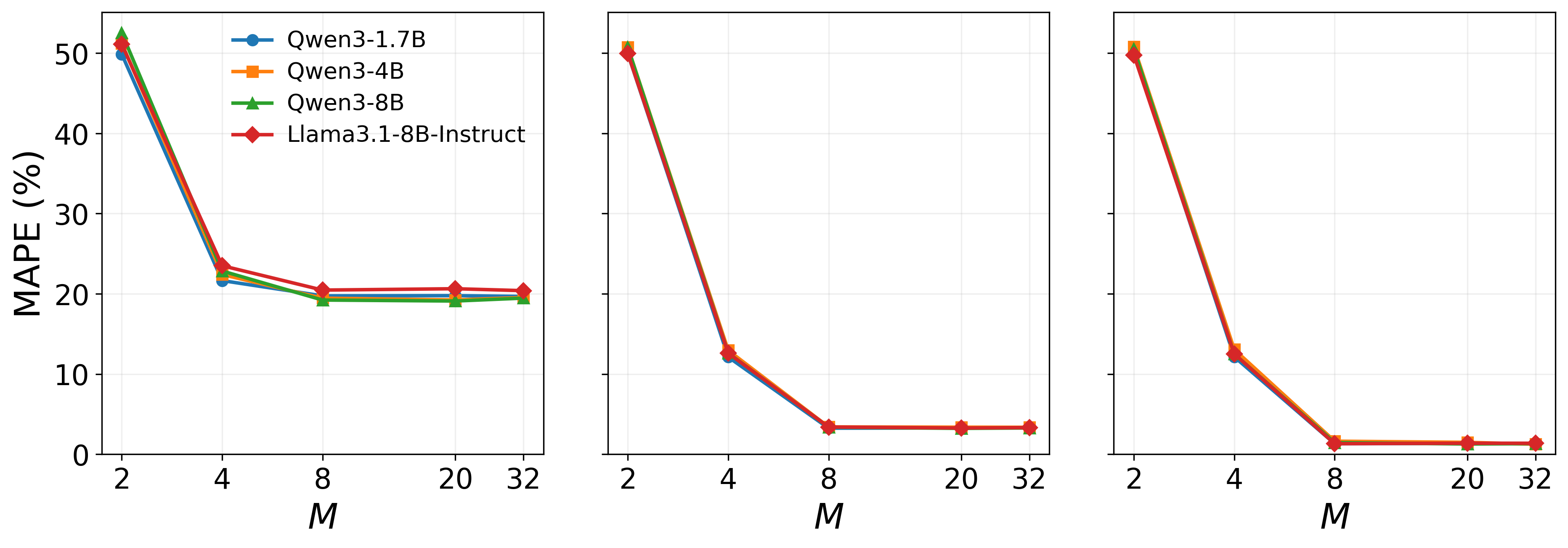}
    \caption{MAPE versus candidate budget $M$ for parallel sampling. Left to right: $\lambda=0.1,0.5,0.9$.}
    \label{fig:parallel-m-ablation}
\end{figure}

Figures~\ref{fig:rejection-m-ablation} and~\ref{fig:parallel-m-ablation} show that both samplers benefit substantially from increasing the candidate budget from $M=2$ to $M=8$, with little further improvement at $M=20$ or $32$. A small budget limits the desired probability of returning a green sequence, as the rejection sampler may reach the cap before encountering a green sequence, and the parallel sampler may have too few green candidates to realize the nominal separation. As a result, a very small $M$ will introduce downward bias.

\subsection{Effect of prompt diversity}
\begin{figure}[p]
    \centering
    \includegraphics[width=1\linewidth]{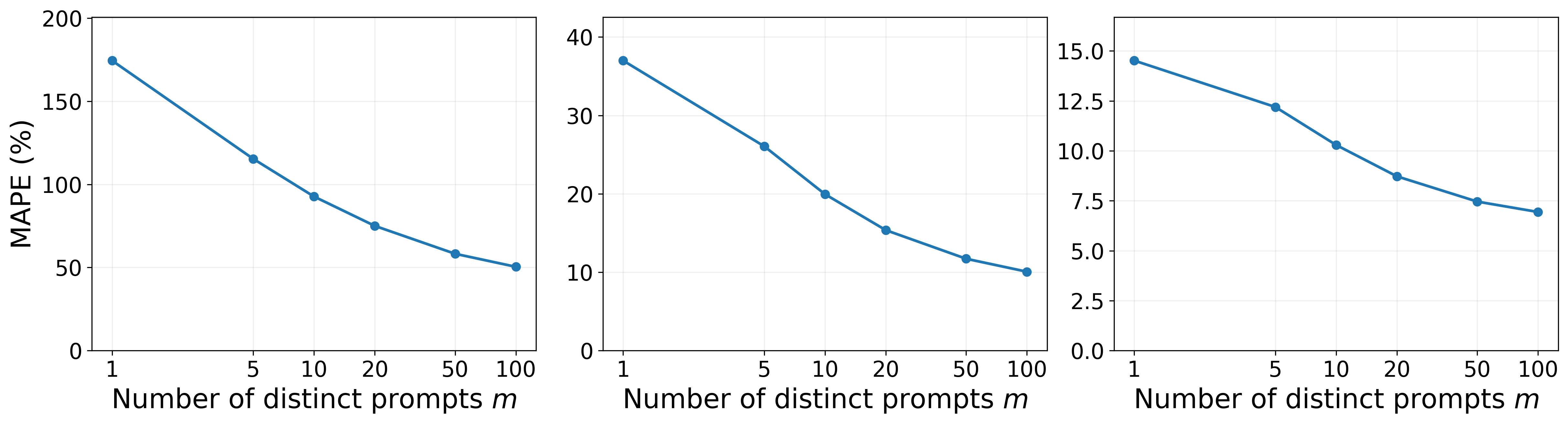}
    \caption{MAPE versus prompt count for Qwen3-4B, with $n=500$ and $T=20$. Left to right: $\lambda=0.1,0.5,0.9$.}
    \label{fig:prompt-diversity}
\end{figure}

We fix the mixture sample size at $n=500$ and vary the number of distinct C4 prompts in the mixture over $\{1,5,10,20,50,100\}$, allocating an equal number of responses to each prompt. We use Qwen3-4B at temperature 1 with $T=20$, rejection RGS with $\eps=2$, $\pi_0=0.5$, and $M=20$. Figure~\ref{fig:prompt-diversity} shows that increasing prompt diversity consistently reduces estimation error when the total sample size is fixed, consistent with theoretical findings in Section \ref{sec:estimation}. 

\clearpage
\section{Examples of RGS-watermarked text}
\label{sec:examples}

\textbf{Setting.} We present example responses generated by the RGS watermark for five C4 prompts with Llama-3.1-8B-Instruct. We use the rejection sampler with $\eps=2$, $\pi_0=0.5$, $M=20$ and temperature $1$. Each prompt is supplied as raw text without a chat
template. The four completions presented for each prompt share the same key, with a different key used for each prompt. We display only the first 50 tokens of the completions. 
\par\medskip
\begin{tcolorbox}[enhanced,breakable,colback=gray!3,colframe=gray!65,
  title={C4 prompts with four RGS-selected completions each},fonttitle=\bfseries,
  fontupper=\small,boxrule=0.5pt,arc=2pt]
\textbf{Prompt 0.} Cluster comprises IBM's Opteron-based eServer 325 server and systems management software and storage devices that can run Linux and Windows operating systems\par\smallskip
\textbf{1.} . IBM has introduced a high-performance computing cluster, called eServer Cluster 1600, which includes the Linux operating system and the Viuke dual-core processor, a x86 chip from Sun Microsystems technology provider: Advanced Micro Devices (AMD). In this\par\smallskip
\textbf{2.} . IBM's "open" cluster, which comprises the eServer 325 (beta), IBM Systems Director software, and several storage subsystems, is generating interest among customers wanting a high-performance computing setup that accommodates a mix of operating systems, including\par\smallskip
\textbf{3.} . Oracle has announced Oracle9i Release 2 Standard Edition, which includes a standard edition of Oracle9i Database Server, Oracle9i Application Server, Oracle Portal and Applications. Release 2 adds improved high availability, the ability to integrate business processes\par\smallskip
\textbf{4.} . IBM is pushing the M@L cluster high-end servers into swift and fast implementation paths, highlighting Linux and Windows combinations. The recommendation is for IBM eServer cluster to ease and accelerate Highly-Scalable and optimized cluster solutions for employment by businesses\par\smallskip
\par\medskip\hrule\medskip
\textbf{Prompt 1.} Belying expectations, Prasar Bharti has earned only Rs 58.19 crore (Rs 581.9 million) as revenue during the\par\smallskip
\textbf{1.} 2006-07 financial year. The figure is significantly lower than the Rs 750 crore (Rs 7.5 billion) expected by the organisation as an immediate step towards achieving its annual targets. The shortfall as estimated appears to be mainly due\par\smallskip
\textbf{2.} first quarter of FY2023-24, till September 30th, which is just 38\% of the total collection of Rs147.120 crore (Rs 1.4712 billion), projected for the entire year. The numbers were shared\par\smallskip
\textbf{3.} current fiscal as against the budgeted estimate of Rs 1,061 crore (Rs 10.61 billion). According to the Prasar Bharti's Annual Report 2010-11, the public sector undertaking (PSU) has failed to\par\smallskip
\textbf{4.} April-June quarter, less than Rs 172 crore (Rs 1.72 billion) in the same period during the previous year. According to the revenue figures released by the ministry of information and broadcasting, the significant drop in revenue is partly due\par\smallskip
\par\medskip\hrule\medskip
\textbf{Prompt 2.} Marketers are employed in the public sector as well as the private sector. Most people think of marketing as a strictly private sector activity, but\par\smallskip
\textbf{1.} in reality, many public sector jobs also involve marketing. Marketing is the functional process for which teams are tasked with the planning, creation, and distribution of ideas, goods, services, and values to develop cohort, create Brand awareness, and to achieve [\par\smallskip
\textbf{2.} public sector employees also market products and services. Additionally, many marketers work in non-profit organizations and professional associations. After obtaining employment, they may assume roles which help to bring their product or services to their customers, then, develop and execute marketing strategies aimed\par\smallskip
\textbf{3.} many government services and public institutions employ marketing professionals to convey information to the public. Overall, there are a wide range of careers to be found in both public and private sectors. The common term "marketer" generally refers to any marketing professional. Other\par\smallskip
\textbf{4.} it is also a vital part in the public sector. While government and non-profit organisations may not have sales objectives in the same way as private companies, the need for customer awareness and a positive image translates to a need for effective marketing. In the public\par\smallskip
\par\medskip\hrule\medskip
\textbf{Prompt 3.} MIDDLETOWN, NJ - The Middletown Township Public Library and the Township of Middletown are pleased to announce the\par\smallskip
\textbf{1.} launch of "Park ...Then", an innovative new program that combines public fitness classes, environmental education, and relaxed social gatherings at several of the Township's parks. The program patterned after the successful "Park Fit" program, will feature a series of\par\smallskip
\textbf{2.} launch of a "Fugitive" online public record search tool for the township. The new tool, which is available on the Township of Middletown Municipal Government website, allows users to search for fugitives based on name, race, sex\par\smallskip
\textbf{3.} celebration of their second year of progress in making Middletown a more inclusive and welcoming community. Over the past year, the Township has made a concerted effort to engage in planning and community outreach to identify goals and objectives for inclusion and diversity, with\par\smallskip
\textbf{4.} Middletown-Sayreville-Holmdel-Matawan community wide resource fair, taking place on September 21, from 10:00 a.m. to 2:00 p.m. at the Middletown Township\par\smallskip
\par\medskip\hrule\medskip
\textbf{Prompt 4.} An eagerly awaited report from the World Health Organization (WHO) states that processed meats such as bacon and sausages cause cancer, and red meat\par\smallskip
\textbf{1.} , such as beef, may be carcinogenic to humans. The report, from the WHO’s International Agency for Research on Cancer (IARC), recognizes that a clear link between processed meat consumption and cancer has been demonstrated as a causal link. A class\par\smallskip
\textbf{2.} increases cancer risk. Photo:Getty Australia's biggest meal kit delivery service, Marley Spoon, says it is taking the World Health Organization's recent cancer warnings on processed and red meat seriously. The WHO says that eating processed meat, such as bacon\par\smallskip
\textbf{3.} may also carry a cancer risk. A team of 22 experts from 10 countries conducted a review of more than 800 scientific studies on the relationship between diet and cancer. The report’s findings are as follows: –Consuming 50 grams of\par\smallskip
\textbf{4.} consumption is also linked to an increased risk of cancer. The WHO's International Agency for Research on Cancer (IARC) classified processed meat as "carcinogenic to humans", group 1. This is the highest level of classification, assigned to substances\par\smallskip
\end{tcolorbox}

\end{document}